\documentclass[a4paper,12pt]{article}
\usepackage[T1]{fontenc}
\usepackage{fullpage}
\usepackage{graphicx}
\usepackage{color}
\title{Interactions between financial and environmental networks in OECD countries}
\author{\small{\textit{Franco Ruzzenenti $^{1,3,\star}$, Andreas Joseph $^{2}$\footnote{This paper does not represent the views of the European Central Bank (ECB).
The views expressed are those of the authors alone and do not necessarily reflect those of the ECB.}, Elisa Ticci $^{1}$, Pietro Vozzella $^{4}$ and Giampaolo Gabbi $^{4}$} }}
\date{}
\begin{document}
\maketitle
\begin{center}
\begin{small}
\textit{$^{1}$ Department of Economics and Statistics, University of Siena, Via S.Francesco 1, IT-53100 Siena, Italy\\
$^{2}$ European Central Bank, Kaiserstrasse 29, 60311 Frankfurt am Main, Germany\\
$^{3}$ Department of Biotechnology, Chemistry and Pharmacy, University of Siena, Via Aldo Moro 1, IT-53100 Siena, Italy\\
$^{4}$ Department of Business and Law, University of Siena, Via S.Francesco 1, IT-53100 Siena, Italy\\}

\begin{footnotesize}$^{\star}$ Corresponding author: ruzzenenti@gmail.com\end{footnotesize}

\end{small}
\end{center}

\begin{center}
\section*{Abstract}
\end{center}

\begin{small}
We analyse a multiplex of networks between OECD countries during the decade 2002-2010, which consists of five financial layers, given by foreign direct investment, equity securities, short-term, long-term and total debt securities, and five environmental layers, given by emissions of $NO_{x}$, $PM10$   $SO_{2}$, $CO_{2}$ $equivalent$ and the water footprint associated with international trade. We present a new measure of cross-layer correlations between flows in different layers based on reciprocity. For the assessment of results, we implement a null model for this measure based on the exponential random graph theory. We find that short-term financial flows are more correlated with environmental flows than long-term investments. Moreover, the correlations between reverse financial and environmental flows (i.e. flows of different layers going in opposite directions) are generally stronger than correlations between synergic flows (flows going in the same direction). This suggests a trade-off between financial and environmental layers, where, more financialised countries display higher correlations between outgoing financial flows and incoming environmental flows from lower financialised countries, which could have important policy implications. Five countries are identified as hubs in this finance-environment multiplex: The United States, France, Germany, Belgium-Luxembourg and the United Kingdom.
\end{small}

\begin{flushleft}
Keywords: Finance, Environment, International Trade, Network Theory, Multiplex
\end{flushleft}

\section{Introduction\label{sec_intro}}

The analysis of networks in financial markets has received a growing attention in recent years \cite{catanzaro,stiglitz, battiston,squartinibank,battiston2}. However, most studies to date have focused on closed systems, such as inter-bank markets \cite{iori,delpini}, ownership networks \cite{garlaschellimarket} or networks of directors \cite{battiston1}.
Recently network analysis has been applied to cross-border portfolio investment flows showing that network measurements can be used as indicators of structural robustness of financial systems \cite{andreas}. Research has also been further developed to couple financial networks with real economic networks (cross-border trades, or, the International Trading Network) into an all-inclusive approach \cite{andreas1,andreas2}.  Departing from this latter line of research, the present analysis aims to develop a comparative analysis of the international financial system with the  environmental loads carried by goods in the International Trade Network(ITN) within OECD countries.

Setting off from this latter line of research, the present analysis aims to develop a comparative analysis of the international financial system in conjunction with environmental loads carried by goods in the ITN within OECD countries. The research conducted in this paper relates to two main strands of the literature: the debate on the interactions between financial and bilateral trade flows \cite{blonigen,coeurdacier,aviat} 
and that of the nexus between environmental and investment flows, such as foreign direct investment (FDI)\cite{albuquerque,aizenman,aizenman2}. Our analysis concentrates on correlation and reciprocity structure \footnote{The reciprocity is the share of trade that is mutually exchanged among all nodes in a network}  of relations between finance and the environmental content of trades\footnote{The environmental content of trades refers to the emissions released, directly and indirectly, locally or globally, by the exporter to produce a certain amount of exported good}.
 
\subsection{Interaction between finance and real economy: state of art.}
The existence of reciprocal interaction between financial flows (particularly long term ones, such as FDI) and trade between countries is well established in the literature\cite{blonigen}. According to the model proposed by Coeurdacier \cite{coeurdacier}, trade openness, namely the exposure of domestic firms to increasing international competition, might foster the acquisition of  foreign firms equities as a hedging strategy \footnote{Hedging: making an investment to reduce the risk of adverse price movements.}. Consequently, bilateral equity holdings and commercial imports are expected to be positively correlated. Trade relations could also lead to a reduction in borrowing costs which, in turn, stimulates investments. Information asymmetries might provide another explanation: transactions in international trade facilitate information flows between trading partners which, in turn, lowers uncertainties for international financial transactions and vice versa \cite{aviat}. A growing body of empirical evidence corroborates the hypothesis of the complementarity between trade and FDI \cite{albuquerque,aizenman,aizenman2}. The empirical literature has also investigated the correlation between trade and portfolio investments. Aviat and Coeurdacie \cite{aviat}, exploring the geography of trade in goods and asset holdings, for instance, find that the causality between bilateral asset holdings and commercial trade strongly progresses in both ways. Lane and Milesi-Ferretti \cite{lane}, using data on international portfolio positions, show that there is a strong correlation between bilateral equity holdings and bilateral trade in goods and services, however the question should be asked if this connection between trade and financial flows has environmental implications and whether highly financialised countries have incentives to re-allocate industrial production to less developed countries to reduce their domestic pollution.

Recent contributions on the "pollution haven effect" suggest that stringency of environmental regulation affects country competitiveness reducing net exports, increasing net imports and affecting firms' location choice, and consequently FDI. Additionally analyses by Aichele and Felbermayr \cite{aichele} reveal that the Kyoto Protocol affects trade flows by significantly increasing committed countries'  embodied carbon imports from non-committed countries and the emission intensity of their imports. Less attention is paid to the role for portfolio investment as a channel for high financialised countries to re-allocate the industrial production into other countries to reduce their domestic pollution.
In our paper, we investigate not only the link between FDI and trade of goods but also the correlation between other financial flows, such as equity and bond securities, which are characterized by a shorter time horizon than FDIs. The research question we want to address is: what role do the long and short term financial flows plays with respect to the direction and intensity of international trade and their expected environmental impact, in terms of different pollution variables?
We find that embodied environmental content of imports have, on average, stronger and more stable correlation and reciprocity with financial outflows, either as FDI and cross-borders portfolio flows, than with inward financial flows. Moreover, this pattern is more accentuated for highly financialised countries. In particular, in the multiplex network of cross-border financial investments and of environmental flows embodied in trade movements, four (France, Germany, USA and UK) out of five hubs are net importers of environmental load.  These results cannot definitively detect a causal link but are consistent with the notion that financial markets help the most financialised countries export capitals in exchange of environment load's displacement. 

The rest of the paper is organized as follows. We first present the used data sources and data content and we briefly present the indexes of correlation and reciprocity calculated in the analysis. Then, we discuss the main results and their economic implications. The last section concludes. Supplementary materials and all details of the analyses are reported in the SM.

\section{Multiplex finance-environment: analysis and results}

\subsection{Data description and sources}
Our network analysis combines various data sources for measures of trade and bilateral financial flows and positions between OECD countries over the 2002-2010 period. More precisely, we consider ten networks where nodes are given by 33 OECD countries and where edges are represented by aggregated/effective cross-border flows of financial investments or goods (See the SI for a detailed list of nodes and flows). Financial flows comprise bilateral Foreign Direct Investment flows (FDI) and portfolio capital flows. FDI data are reported in current USD millions and come from OECD statistics. Portfolio flows are also measured in current USD millions and data are drawn from the Coordinated Portfolio Investment Survey (CPIS) carried out by IMF \cite{lane}. The CPIS records data on the bilateral composition of year-end portfolio holdings (long-term debt, short-term debt and equity portfolio assets) for over seventy reporting/source countries \emph{vis-a-vis} over 200 destination countries, however it does not provide information on portfolio flows. For this reason, all portfolio flows in this paper are measured as the difference between consecutive positions:  an increase (decrease) in debt securities issued by country $j$ and held by residents in country $i$ is recorded as a capital movement from country j to country $i$ (from $i$ to $j$) \cite{andreas}. 

Earlier studies \cite{lane,portfolio2} have already underscored the main limitations of the CPIS in terms of incompleteness, lack of strong data consistence, risk of under-reporting and problematic treatment of intermediated holdings in financial centers. However, the inclusion of portfolio flows, in addition to FDI, allows us to track trends and features of financial investment decisions that are based on different time-horizons and driven by different economic motives. Moreover, we try to contain some of these problems by focusing the analysis on OECD countries which are more likely to report comprehensive information. We therefore select a set of countries which are expected to ensure the best \emph{trade-off} between data quality and geographical and time coverage. This, however, implies that the interpretation of our results should take into account that large players, such as China, and important resource-rich countries, such as the Gulf States, are excluded from the analysis. As for trade networks, we concentrate on environmental flows which are, directly and indirectly, generated by trade flows. The estimation of the environmental footprint of trade flows draws from the Eora global Multi-Region, Input Output (MRIO) database \cite{eora1,eora2}. Eora elaborates a time series of environmentally extended input-output (IO) tables for 187 countries based on the UN System of National Accounts (SNA), UN COMTRADE, Eurostat, IDE/JETRO, and several national IO tables with matching environmental and social satellite accounts. We refer to this database to obtain footprint embodied in bilateral import and export flows in terms of five environmental dimensions: emissions of $(1)$ $NO_{x}$ (in gigagrams, Gg), $(2)$ $PM10$ (Gg), $(3)$ $SO_{2}$(Gg), $(4)$ total $CO_{2}$ $equivalent$ emissions (Gg) $(5)$ and water footprint (in m3).
\subsection{Method and analysis}
The five financial networks analysed in this study comprise the networks of bilateral FDI, portfolio investments in short and long term security debt (SD and LD), in equities as well as in total security debts (SD+LD).. Environmental and financial networks share the same set of nodes, namely the 33 OECD countries. The entanglement of their relationships gives rise to a \emph{multiplex} of interacting networks \cite{DiegoMulti,Bianconi1,Bianconi2}.
Henceforth, every network will be defined as a \emph{layer} of this financial-environmental multiplex. 

\begin{figure}[ht]
\begin{center}
\centerline{
\includegraphics[width=.4\textwidth]{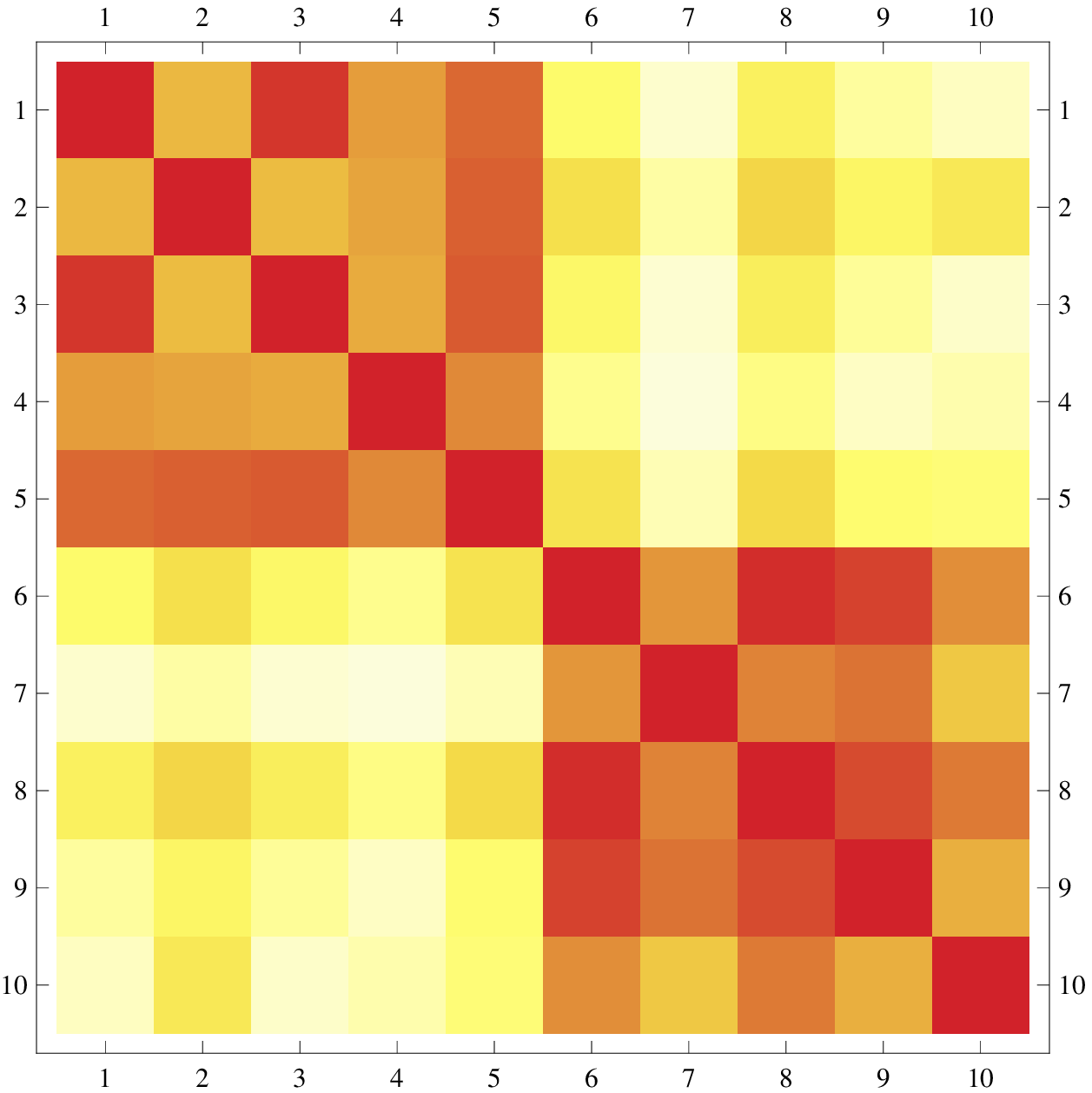}
\includegraphics[width=.4\textwidth]{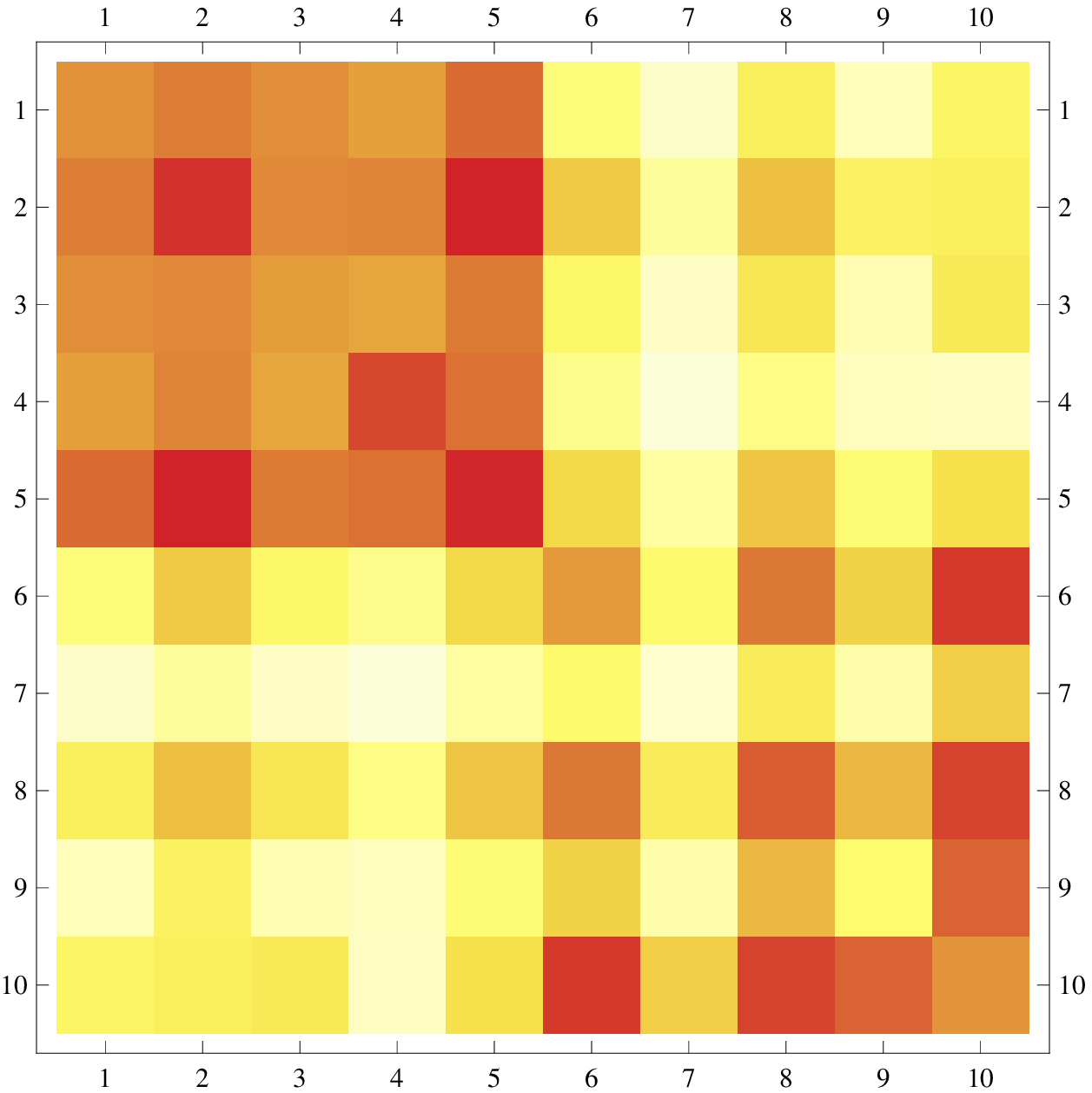}}
\caption{\textbf{Correlations temperature maps} between multiplex layers, synergic (left panel) and reverse flows (right panel), 2002-2010. Shade of colours indicate increasing correlation between couple of layers, from light yellow to dark brown. Layers $1-5$: FDI, Equity, Short-term Debts, Long-term Debts and Total Debts. Layers $6-10$: $NO_{x}$, $PM10$, $(3)$ $SO_{2}$, $CO_{2}$ $equivalent$ and water footprint. According to the Pearson index equity is the financial layer most correlated to the environmental layers, followed by TD (see SI, Table \ref{tab3}). $NO_{x}$, $SO_{2}$ and water are the most correlated layers with the equity layer and most of the other financial layers. The first two polluters are linked to combustion and may hint to a nexus with the heavy industry. The footprint of water is high in the energy intensive and agricultural sectors, and in some manufactures.
}\label{fig1}
\end{center}
\end{figure}

To start with, we investigate the \emph{spatial} correlation among layers by applying a previous method based on the Pearson correlation coefficient \cite{DiegoMulti}. 
Figure \ref{fig1} shows the temperature map (colors from white to dark red indicates an increasing, positive correlation) for the five financial layers (1 to 5) and the five environmental layers (6 to 10) \footnote{Interestingly, but not surprisingly, correlations are all positive: this is due to the fact that the topology in every layer of the multiplex is similar. Hubs of the financial system are also hubs of the ITN. See Table 1 and 2 in SI}. Correlations are averaged over the period under investigation (2002-2010). The first matrix shows correlations \footnote{Correlation matrices for each year are reported in the SI.} between flows going in the same direction (synergic flows) and the second shows correlations between flows going in opposite direction (reverse flows). Interestingly, the most correlated financial layers to the environmental ones are \emph{equities} (layer2) and \emph{total debts} (layer 5). All the environmental layers, are more correlated to $equities$ (with the exception of reverse flows of water, layer 10, that is correlated more with SD, layer2). The second most correlated financial layer to the environment is \emph{TD}. Among the environmental layers, $SO_{2}$ (layer 8) is always the most correlated to any financial layer. This may suggest that financial layers tend to be more correlated to manufacturing, metallurgy and industry in general that relies heavily on the combustion of raw fossil fuels, but more research is needed on this aspect. The second most correlated is $NO_{x}$ (layer 6), further suggesting a link to combustion and, foremost, to low-efficiency combustion. This may hint at a specific topology in correlation that will be addressed further in a subsequent section of this paper.
A second notable result of our analysis is that, in most cases (15 out of 25, see Table \ref{tab3} and \ref{tab4} in SI) and particularly for the most correlated layers (equity, TD, $SO_{2}$ and $NO_{x}$), \emph{reverse} flows display higher correlation to the environment than synergic flows, indicating that among OECD countries financial flows tend to be associated to environmental flows going in the \emph{opposite direction}, more often than flows going in the \emph{same direction}. Furthermore, reverse correlations are more stable than synergic correlations (see SI for further details). Reverse correlations among flows suggest that different layers tend to reciprocate.
We measure the cross-product reciprocity between layer $F$ and $E$ and country $i$ and $j$, normalized on country's exports ($Exp$) \cite{Wrecip}: 

\begin{equation}
r_{ij}^{FE}=\frac{X_{ij}^{F}X_{ji}^{E}}{\sum_{j}X_{ij}^{F}\sum_{j}X_{ji}^{E}}
\label{localr}
\end{equation}

\begin{figure}[ht]
\begin{center}
\centerline{\includegraphics[width=.4\textwidth]{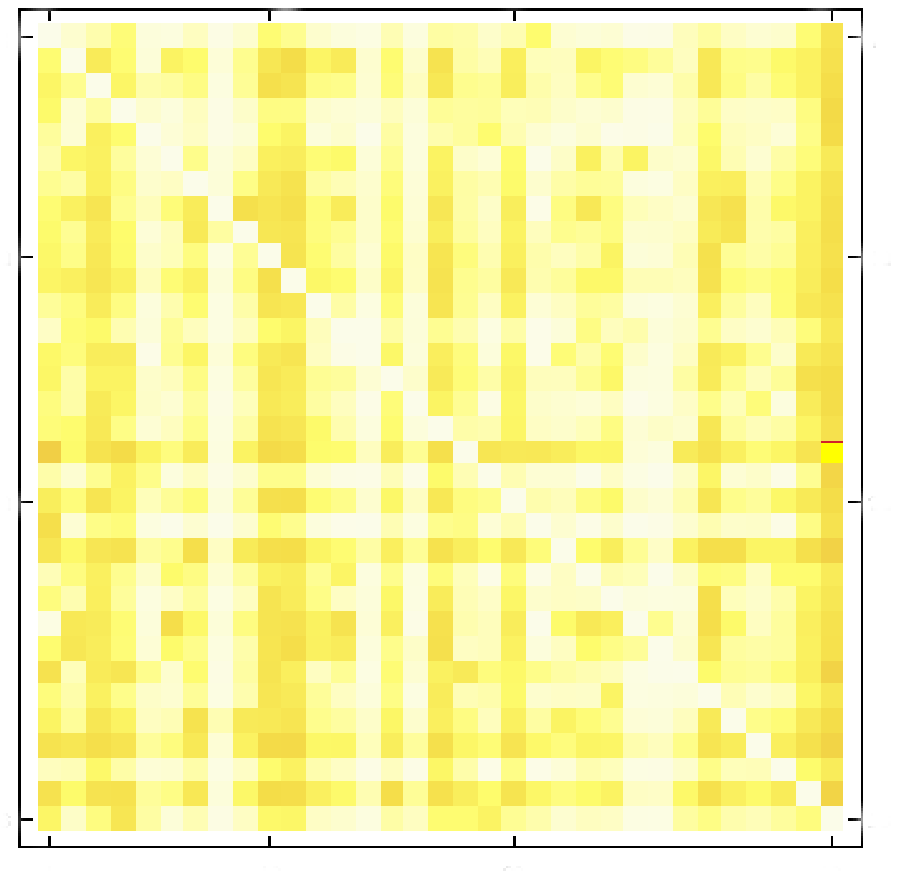}
\includegraphics[width=.4\textwidth]{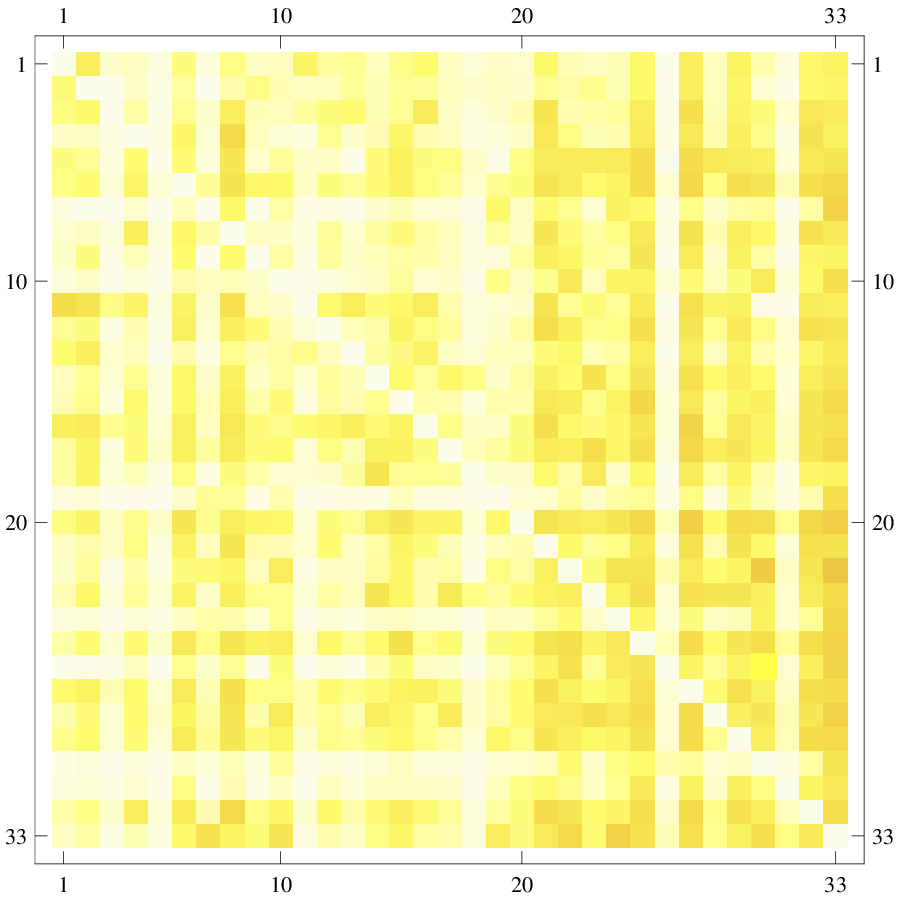}}
\caption{Temperature maps of \textbf{local reciprocity} between financial outflows and environmental inflows (normalized over exports), country panel: countries in alphabetical order, left to right (left panel) and countries ordered for increasing financialisation (right panel). Years 2002-2010. From left to right the emerging gradient of yellow indicates that there is a topology in the reciprocity structure that is proportional to the financialization of countries.}\label{fig2}
\end{center}
\end{figure}

Figure \ref{fig2} shows the cross-product reciprocity between financial layers on aggregate (the average of the four financial layers) and environmental layers on aggregate (the average of the five environmental layers). In the first matrix (Figure \ref{fig2}, left panel), entries (countries) are reported in alphabetical order, while in the second (Figure \ref{fig2} right panel), countries are ordered by increasing financialisation (measured as value added of the financial sector over the total value added). Shifting the array-order of countries we can see a gradient of increasing reciprocity emerging from the matrix, when going from less to more financialised countries. This result confirms that there is a specific topology in the financial-environment correlation network that mirrors the level of countries' financialisaton. The top-right block of the matrix (Figure \ref{fig2} right panel) shows an average higher reciprocity than the bottom-left block (see Table \ref{tab4} in SI). This asymmetry is evident when looking at the different degrees of yellows in the two regions, indicating a dominant pattern of reciprocity. Exports of finance from the most financialised countries (entries 19-33) towards the less financialised (entries 1-18) are more correlated to environmental imports than flows from low financialised to high financialised countries. The countries of the low financialised group show higher correlation with the countries of the high financialised group than within themselves, contrary to the the counties of this latter group, which are instead tightly connected among themselves. In other words, it seems that highly financialised countries tend to exchange financial flows with environmental flows with countries that are less financialised, by means, most prominently in terms of equities and TD.

\begin{figure}[ht]
\begin{center}
\centerline{\includegraphics[width=.4\textwidth]{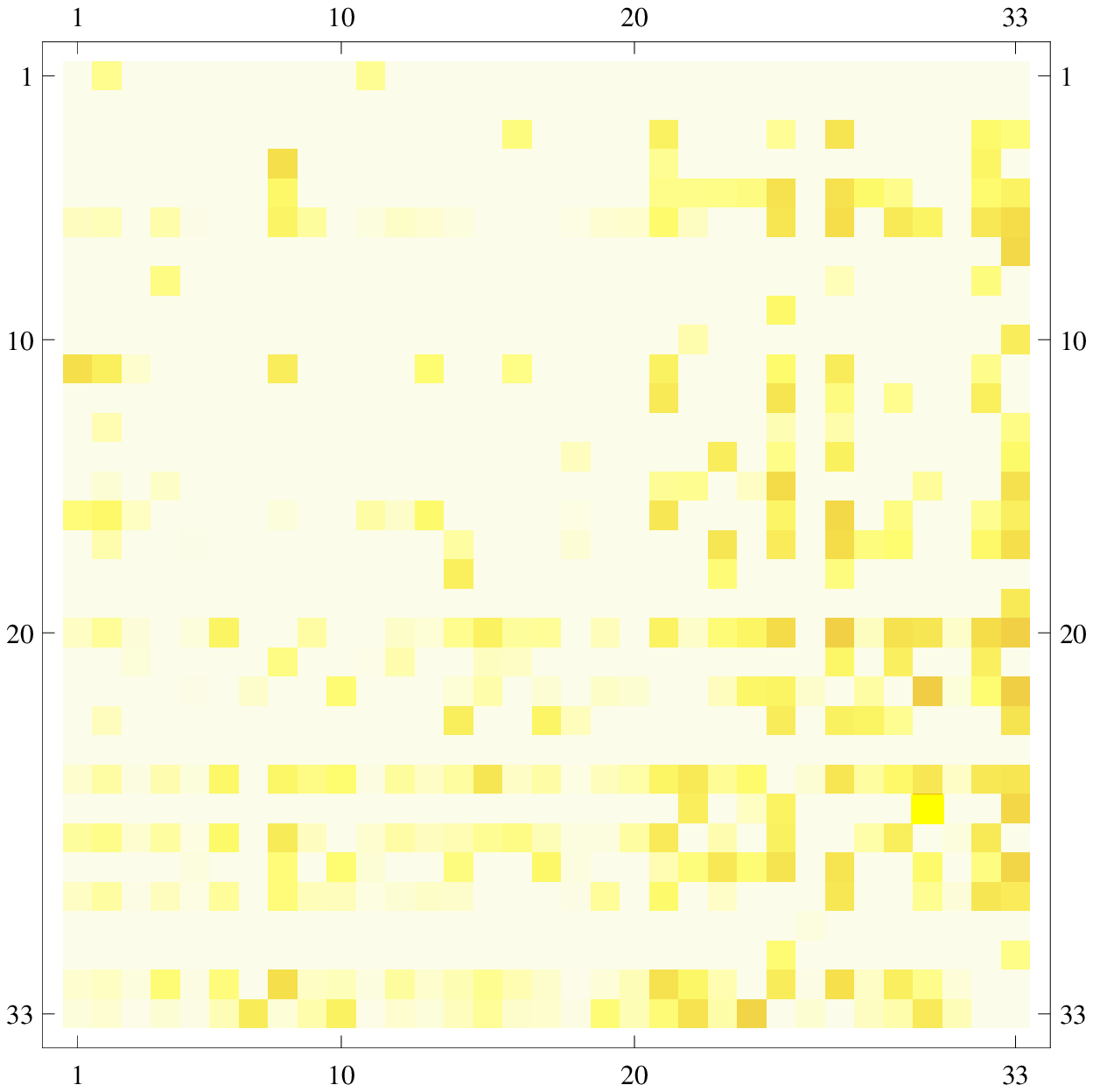}
\includegraphics[width=.6\textwidth]{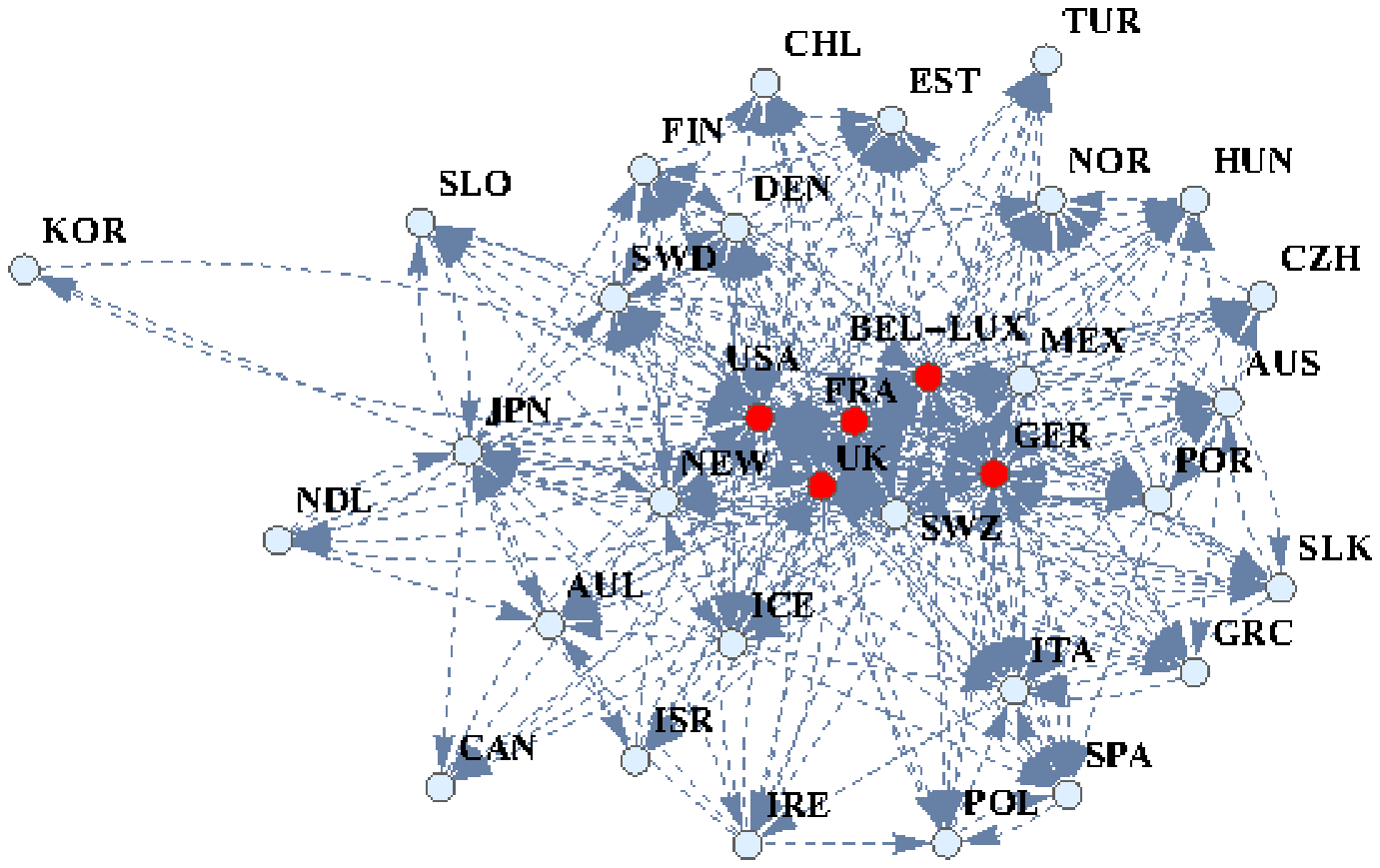}}
\caption{\textbf{Null-model-enhanced local reciprocity} between financial outflows and environmental inflows (local $\rho$); countries ordered by degree of financialisation, left to right (years 2002-2010). Financial flows are outgoing and environmental flows are incoming; exports are normalized over columns (example: entry $w_{ij}$ indicates the share of financial export of country $j$ that is reciprocated by the share of environmental export of country $i$. On the left the temperature map of the reciprocity values that exceed the significance threshold posed by the null model. Wihte dots indicate that there is no significant reciprocity between countries. Yellow dots concentrate in two regions: bottom-right and upper-right quadrants. The first indicate significant reciprocity among higly financialized countries. The second that there is a significant reciprocity from higly financialsed countries to less financialised countries. The The right panel shows the backbone network of the reciprocity structure: an arrow indicate a link of an outgoing financial flow reciprocated by an incoming environmental flow. The five hub-countries are shown with red dots.}\label{fig3}
\end{center}
\end{figure}

Given the prevalence of reverse correlations, we can focus our analysis on reciprocal exchanges between the financial layers and environmental layers. It should be noted that a \emph{multiplex} is a very complex structure \cite{Bianconi1,Bianconi2}. Spurious correlations may rise from overlapping effects driven by two factors: the reciprocity structure within each layer and the topology peculiar to each layer. In what follows, we want to disentangle the correlation between each pair of layers from the topology specific to every single layer (with respect to our analysis, reciprocity between layers can be considered as a different form of correlation). By utilising a methodology recently developed in \cite{diegomulti1,diegomulti2}, we are able to provide a measure of correlation between layers for reverse flows, on both a global ($\rho$ correlations) and local scale (equation \ref{localrho}) that incorporates a null model and thereby clearing spurious effects from our analysis:
\begin{equation}
\rho_{ij}^{FE}=\frac{r_{ij}^{FE}-\langle r_{ij}^{EE}\rangle}{1-\langle r_{ij}^{FE}\rangle}
\label{localrho}
\end{equation} 
This measure signals when reciprocity exceeds the expected reciprocity trivially produced by the null model. The chosen null model is based on an exponential randomization that preserves export, import and reciprocated trade flows for each country pair, which are set as constrains into the model \cite{Wrecip}. 
On the one hand, this null model enables us to test previous results and, on the other hand, to extract the backbone of significant correlations among countries. According to the cross-product reciprocity and the statistical validation, results of the Pearson index hold: equities and TD are more correlated to environmental flows than FDI and reverse correlation tends to prevail over synergic correlation. We can further test the structure of the reciprocal relationship among OECD countries between financial outgoing flows and environmental incoming flows with the null model based on Exponential Random Graph (see SI). The matrix of Figure \ref{fig3} shows the reciprocal exchanges between financial layers and environmental layers exceeding the single-layer reciprocity level set by the null model. The degree of yellow signals positive values of $\rho$ whereas the white dots indicates that there is no significant correlation (equation \ref{localrho}). The top-right block shows, on average, stronger correlations than the bottom-left, pointing to a statistically significant connection between financial outgoing flows reciprocated by incoming environmental flows in the region of the highly financialised countries with respect to the lowest financialised countries (Table \ref{tab3} in SI).  

\begin{figure}[ht]
\begin{center}
\centerline{\includegraphics[width=.9\textwidth]{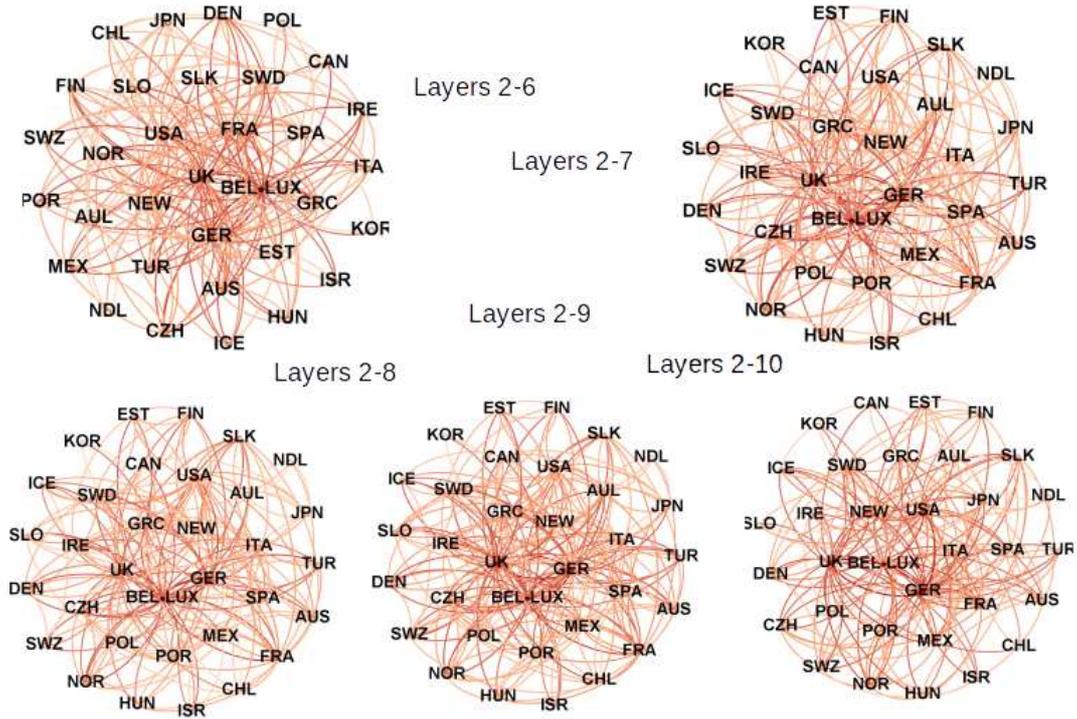}
}
\caption{Backbone of the links of null-model-enhanced local reciprocity, between the \textbf{equity layer} and the five environmental layers, for the year 2010. From left to right: $NO_{x}$, $PM10$, $(3)$ $SO_{2}$, $CO_{2}$ $equivalent$ and water footprint. Increasing dark red indicate an increasing out-degree of the node. The hubs are pleaced in the core of the cloud. The reciprocity analysis confirms that equity is mostly reciprocated with $NO_{x}$ and $SO_{2}$, suggesting a link with the industrial sector.}\label{fig4}
\end{center}
\end{figure}

The yellow dots in Figure \ref{fig3} can be converted into directed links in order to depict the topology of reciprocal relations among OECD countries. In what follows, links map the network backbone of positive correlations exceeding the level of significance posed by the null model. The arrows in the right panel of Figure \ref{fig3} stand for a statistically significant outgoing financial flow reciprocated by an incoming environmental flow. For almost all layer pairs and for the financial-environmental layers in aggregate, the backbone of the correlation network highlights the central role of five distinctive nodes: USA, France, Germany, Bel-Lux and UK. Figure \ref{fig4} show the case of the equity layer compared to the five environmental layers.
Those five countries display double the number of links compared to the other OECD countries (Figure \ref{S3 Fig}). This observation mirrors the central role these countries take in both financial markets and international trade. Contrary to most of the developed and financialised
countries, USA, France, Germany,and UK, are net importers of environmental flows, with the remarkable exception of Bel-Lux (Figure \ref{S4 Fig}). At the same time, but not surprisingly, these five countries are also net importers of financial flows (Figure \ref{S5 Fig}). It is noteworthy that these five countries \footnote{With the exception of Germany, which is net exporter of value and net importer of mass.} are net importers in terms the value and the mass of international trade of goods. Industrialized countries transform row materials into composite goods. This explains their negative mass imbalance. Goods that are positioned higher in the value chain have arguably a lesser content of mass and an higher environmental footprint by unit of value.

\section{Conclusions}
In this paper, we have focused on the linkage between financial and environmental movements which are generated by cross-border investment activities and the trade of goods. In contrast to previous studies, we have compared the correlation between environmental and portfolio investment flows, including FDI as well as IPI. We have found that most environmental indicators, apart from water, are highly correlated with short term financial flows (which might also be more speculative in nature) than FDIs, which are mainly long-term and stable investments in the real economy. 
This result appears to be coherent with the hypothesis that short-termism is not only linked to the financial system's stability but also to the real economy and environmental sustainability. A second finding is that reverse flows, particularly short term equity trades, show higher correlation with environmental flows than synergic flows, indicating that, among OECD countries, financial flows tend to be associated with environmental flows going in the opposite direction. Reverse correlations among flows suggest that different layers tend to reciprocate. Furthermore, the dominant pattern in reverse correlations between finance and environment is seen to be directed from more financialised to less financialised countries. The conclusion from this is that agents in net-importer and highly financilised countries tend to take speculative exposures in stocks traded in less industrialized countries. 
However, not all fiancialised countries exhibit a signicant correlation between financial layers and environmental layers (see Table \ref{tab_OECD} of SI for the complete ranking of OECD countries). The back-bone within this highly complex structure is formed by five countries, which are themselves hubs for international trade and investment, namely: the USA, France, Germany, Bel-Lux and the UK. These five countries are both net importers of finance and net importers of mass. Do these results, concerning the topology of correlations, hold even if we would enlarge the scope of the analysis, including emerging economies and developing countries? What is the causal relationship beyond these correlations? Do financial investments draw environmental load or financial flows follow the channels of trade? These questions remain unanswered because of limitations of the model and uncertainties in the data. Although we cannot derive conclusions on causality, we built a first step in contributing to these areas of research in three main ways: by applying network analysis instruments to study the links between financial flows and trade-related environmental movements (beyond bilateral view), by including portfolio investment flows (beyond FDI), and by evaluating trade flows according to their environmental content rather than in monetary or mass terms (beyond disciplinary separatism). What we get is a picture of the world, albeit reduced to only OECD countries, of immense complexity, in which the interacting systems,  even those seemingly distinct, can not be seen separately.
To say it succinctly, with the stylish and sarcastic words of Oscar Wilde\cite{owilde}:

\begin{quote}
London is too full of fogs and serious people.
Whether the fogs produce the serious people
or whether the serious people produce the fogs,
I don't know.
\end{quote}

\section{Acknowledgements}
We would like to thank Diego Garlaschelli, Tiziano Squartini, Francesco Picciolo and Rossana Mastandrea for their help. Without their contribution this work would have not been possible.

\section*{Supporting Information \label{SI_text}}

What is a \emph{multiplex}? A multiplex is a bundle of networks (layers) that share the \emph{same nodes} but potentially with different flows (edges) within each layer. Each network, or \emph{layer}, can feature relationships of the \emph{same nature} and measured in the same unit or can display a broad variety of connections and, thus, links will be measured in different units \cite{Bianconi1,Bianconi2}. Metrics of our analysis will be of two natures: monetary units (USD) for the financial layers and mass units for the environmental layers (table \ref{tab_layers}). The nodes of the multiplex are 33 OECD countries and will be ordered in every layers' matrix adjacency according to the increasing financialisation of their economy,  expressed by value added accounted by financial intermediation, real estate, renting and business activities as a percentage of total value added in 2000, namely prior to the period under study (Table \ref{tab_OECD}). We analyzed the interactions of the resulting financial-environmental multiplex over a time span of 9 years, from 2002 to 2010.
These multiplex' interactions over 9 years for 10 layers and 33 nodes can be expressed as a tensor $\sigma_{tkij}$ with dimensions ($9x10x33x33$) and the entry of the tensor $w(t)_{ij}^{k}$  expresses the flow of the layer $k$ at time $t$ from country $i$ to country $j$. The total dataset for this analysis will thus has thus 95040 data points. 

\begin{table}
\caption{List of multiplex' layers}
\centering
\begin{tabular}{|l|l|l|}
\hline
\multicolumn{2}{|c|}{\textit{Financial networks}}\\
\hline
FDI & Foreign Direct Investment & 1 \\
Equity & & 2 \\
SD & Short Term debts & 3 \\ 
LD & Long Term debts & 4 \\
TD & Total debts & 5 \\ 
\hline
\multicolumn{2}{|c|}{\textit{Environmenatl networks}}\\
\hline
NOx & Nitrogen oxide & 6 \\
PM10 & Particulate Matter  & 7 \\
SO2 & Sulfur dioxide  & 8 \\
CO2 & Carbon Dioxide & 9 \\
Water& & 10 \\
\hline
\end{tabular}
\label{tab_layers}
\end{table}

\begin{table}
\caption{List of multiplex' nodes}
\centering
\begin{tabular}{|l|l|l|}
\hline
Country & ranking & FIRE/TV \footnote{Share of financial intermediation, real estate, renting and business activities over total value added}\\
\hline
Czech Republic & 1  & 16.2\\
Norway & 2 & 16.9\\
Slovak Republic & 3 & 17.1\\
Poland & 4 & 18.1 \\
Iceland & 5 & 18.9 \\ 
Mexico & 6 & 19.0\\
Korea & 7 & 19.3\\
Spain & 8 & 19.5\\
Turkey & 9 & 19.5 \\
Slovenia & 10& 20.2 \\
Portugal & 11 & 20.3\\
Greece & 12 & 20.6 \\
Hungary & 13 & 20.9\\
Finland & 14 & 21.0\\
Ireland & 15 & 21.3\\
Austria & 16 &21.5\\
Denmark & 17 & 22.3 \\
Estonia & 18 & 22.4\\
Chile & 19 & 23.1\\
Switzerland & 20 & 24.0 \\
Italy & 21 & 24.7\\
Japan & 22 & 24.9\\
Sweden & 23 & 24.9 \\
Canada & 24 & 25.0\\
United Kingdom & 25& 27.0\\	
Netherlands & 26 & 27.3\\
Germany & 27 & 27.5 \\
New Zealand &  28 & 27.8\\
Bel-Lux & 29 & 29.1\\
Australia & 30 &29.1\\
Israel & 31  & 30.5\\
France & 32 & 30.7\\
United States & 33 & 31.7\\
\hline
\end{tabular}
\label{tab_OECD}
\end{table}

\subsection*{A measure of correlation: the Pearson correlation index\label{pearson}}
A first method to investigate layers' correlations in a multiplex framework was proposed by  Garlaschelli and applied to the commodity-specific trades \cite{DiegoMulti}. Garlaschelli extended the Pearson correlation index to the multiplex by averaging over space instead of time. The Pearson correlation index between the layers A and B  (at time $t$) will thus be:

\begin{equation} 
\rho^{AB}_{Syn}\equiv\frac{\sum_{i\neq j}(w^{A}_{ij}-\mu^{A})(w^{B}_{ij}-\mu^{B})}{\sqrt[2]{\sum_{i\ne j}(w^{A}_{ij}-\mu^{A})^{2} (w^{B}_{ij}-\mu^{B})^{2}}}
=\frac{cov_{AB}}{\sigma_{A}\sigma_{B}}.
\label{pearson_syn}
\end{equation}

It is noteworthy that the above definition of correlation can be applied both to links going in the same direction (equation \ref{pearson_syn}) and to links going in opposite directions (equation \ref{pearson_rev}). We call the former \emph{synergic} (syn) correlations and the latter \emph{reverse} (rev) correlation.

\begin{equation} 
\rho^{AB}_{Rev}\equiv\frac{\sum_{i\neq j}(w^{A}_{ij}-\mu^{A})(w^{B}_{ji}-\mu^{B})}{\sqrt[2]{\sum_{i\ne j}(w^{A}_{ij}-\mu^{A})^{2} (w^{B}_{ji}-\mu^{B})^{2}}}
\label{pearson_rev}
\end{equation}

Both kinds of correlations, \emph{syn} and \emph{rev}, show that the multiplex is divided in two separate blocks: the financial block (entries 1 to 5), with a weak inner correlation and the environmental block, with a stronger inner correlation. It should be noted that the diagonal elements of the $syn$ correlation matrix are $1$ by definition, whereas the $rev$ correlations on the diagonal score $1$ only when the layer's original matrix is symmetrical. Therefore, elements on the diagonal of the $rev$ correlation matrix are measures of the symmetry of the single layer matrix. Indeed, this is a hint of the entanglement of reciprocity and correlation in multiplex networks.

Equation \ref{pearson_rev} can be applied to the binary structure of layers, where the weighted entries $w^{k}_{ij}$ of layer $k$ are replaced by the binary entries $a^{k}_{ij}$ that score $1$ when a link exist between node $i$ and $j$ and $0$ otherwise.
For the binary representation of networks, the Pearson correlation matrix is thus:
\begin{equation} 
\rho^{AB}_{b}\equiv\frac{\sum_{i\neq j}(a^{A}_{ij}-\mu^{A})(a^{B}_{ij}-\mu^{B})}{\sqrt[2]{\sum_{i\ne j}(a^{A}_{ij}-\mu^{A})^{2}(a^{B}_{ij}-\mu^{B})^{2}}}
\label{pearson_binary}
\end{equation}

The binary correlation matrix of the multiplex indicate that there are two blocks on the diagonal, the financial block is sparse, measured by the number of connections, and the environmental block that is composed by fully connected layers. The $connectance$ (the ratio of existing links over the number of possible links $N(N-1)$,  given a network with $N$ nodes, equation \ref{r2}) of the environmental networks is $1$ for every layer, whereas in the financial networks it oscillates around $0.5$. It is noteworthy that, in a dense network, mutual links are more likely to occur than in a sparse network and in a fully connected network the reciprocity is \emph{trivially} maximal.

For binary and directed networks, the reciprocity is defined as the fraction of links having a ``partner'' pointing in the opposite direction:

\begin{equation}
r^b\equiv\frac{L^\leftrightarrow}{L}
\label{r}
\end{equation}

\noindent where $L=\sum_{i\ne j}a_{ij}$ and $L^\leftrightarrow=\sum_{i\ne j}a_{ij}a_{ji}$. The above quantity, $r^b$, is not independent on the link density (or connectance) $c\equiv\frac{L}{N(N-1)}=\frac{\sum_{i\ne j}a_{ij}}{N(N-1)}\equiv\bar{a}$: on the contrary, it can be shown that $c$ is the expected value of $r^b$ under the Directed Random Graph Model  \cite{mygrandcanonical,myreciprocity}. 
In the DRG, a directed link is placed with probability $p$ between any two vertices, i.e. $\langle a_{ij}\rangle_{DRG}=p,\:\forall\:i, j$ (with $i\neq j$). This implies

\begin{equation}
\langle r^b\rangle_{DRG}\equiv\frac{\langle L^{\leftrightarrow}\rangle}{\langle L\rangle}=\frac{N(N-1)p^{2}}{N(N-1)p}=p\equiv\frac{L}{N(N-1)}=c
\label{r2}
\end{equation}

\noindent showing that the expected value of $r^b$ coincides with the fundamental parameter of this null model, and hence depends on $L$ and $N$.
In order to assess whether there is an actual tendency in the network to establish mutual links, one should compare the measured $r^b$ with its expected value $\langle r^b\rangle_{DRG}$.
This means that $r^b$ cannot be used to consistently rank networks with different values of $L$ and $N$, because of their different reference values. 

\subsection*{Reciprocity as a correlation coefficient\label{sec_rec}}

A measure of reciprocity based on the Pearson correlation index was proposed in \cite{myreciprocity}. Reciprocity is therefore the Pearson correlation coefficient between the transpose elements of the adjacency matrix  \cite{mysymmetry2}:

\begin{equation} 
\rho^b\equiv\frac{\sum_{i\neq j}(a_{ij}-c)(a_{ji}-c)}{\sum_{i\ne j}(a_{ij}-c)^2}=\frac{r^b-c}{1-c}=\frac{r^b-\langle r^b\rangle_{DRG}}{1-\langle r^b\rangle_{DRG}}.
\label{rho}
\end{equation}

A symmetrical adjacency matrix  represents a network with the highest values of $r^b$ and $\rho$ (both equal to 1, thus regarding a fully connected network), whereas a fully asymmetrical one, with $zero$ values mirroring $unit$ values on opposite sides of the main diagonal (like a triangular matrix), displays the lowest value, being $r^b=0$ and $\rho=-c/(1-c)$) \cite{myreciprocity}.
This meaningful definition of reciprocity automatically discounts density effects, i.e. the expectation value of $r^b$. As a result, consistent rankings and temporal analyses become possible in terms of $\rho$.
However, this definition of reciprocity, mutated from Pearson, only works for single binary networks. What would be a definition of reciprocity for a weighted multiplex? First, we will extend the definition of binary reciprocity to weighted networks. 

\subsection*{From binary to weighted\label{sec_rec2}}

If we follow the binary recipe from left to right, we define the weighted reciprocity as the Pearson correlation coefficient (where, as usual, $\bar{w}=\frac{\sum_{i\neq j}w_{ij}}{N(N-1)}=\frac{W_{tot}}{N(N-1)}$): 

\begin{equation} 
\rho\equiv\frac{\sum_{i\neq j}(w_{ij}-\bar{w})(w_{ji}-\bar{w})}{\sum_{i\ne j}(w_{ij}-\bar{w})^2}=\frac{r-c^w}{1-c^w}
\label{rhow}
\end{equation}

\noindent where, in order to produce a result formally equivalent  to eq.(\ref{rho}), we have defined the weighted analogues of $r$ and $c$ as follows:

\begin{equation} 
r\equiv\frac{\sum_{i\neq j}w_{ij}w_{ji}}{\sum_{i\neq j}w_{ij}^2},\:c^w\equiv\frac{\bar{w}^2}{\sum_{i\neq j}w_{ij}^2/N(N-1)}
\label{rcw}
\end{equation}

\noindent Note that the equivalence $\bar{a}=c$, valid for the binary case, no longer holds: $\bar{w}\neq c^w$. The previous expressions generalize the binary ones and reduce them when substituting  $a_{ij}$ for $w_{ij}$. 
In \cite{Wrecip}, we proposed a different measure:
\begin{equation}
r\equiv\frac{W^{\leftrightarrow}}{W}=\frac{\sum_{i\neq j}\min[w_{ij},\:w_{ji}]}{\sum_{i\neq j}w_{ij}}.
\label{rmin}
\end{equation}
Indeed, the few attempts that have been made so far in order to characterize the reciprocity of weighted networks \cite{jari,fagiolo,achen,faloutsos} are all based on measures of correlation or symmetry between mutual weights. In \cite{Wrecip} it has been shown that a  measure of reciprocity based on summation of the $minimum$ values of links' weight between every dyad (couple of nodes) is more reliable and promising .

\subsection*{A correlation measure for a multiplex: cross product reciprocity and multiplexity\label{sec_rho}}
A multiplex can be featured by homogeneous layers -layers whose nature of trades and thereby unit of account, is the same throughout, or heterogeneous -multiplex composed by layers of different nature and units. The multiplex of the present analysis belongs to the latter case. However, does the definition of reciprocity in equation (\ref{rmin}) apply to both cases?
Although the definition of reciprocity based on $minimum$ (equation \ref{rmin}) is a more refined measure, as provided in  \cite{Wrecip}, it has one major flaw when applied to a multiplex: it is sensible to the scale of layers:
\begin{equation}
r^{AB}=\frac{2\sum_{i\neq j}\min[w^{A}_{ij},\:w^{B}_{ji}]}{\sum_{i\neq j}w^{A}_{ij}+\sum_{i\neq j}w^{B}_{ij}}.
\end{equation}
This problem becomes detrimental when dealing with layers of very different nature and, thus, units. Setting the informative level of reciprocity, based on the minimum amount exchanged, would draw to the arbitrary decision of the scale of the unit. Therefore, we will explore a  measure for the multiplex (cross) reciprocity based on equation \ref{rcw}:
\begin{equation}
r^{AB}=\frac{\sum_{i\neq j}w_{ij}^{A}w_{ji}^{B}}{\sum_{i\neq j}w_{ij}^{A}\sum_{i\neq j}w_{ij}^{B}}=\frac{\sum_{i\neq j}w_{ij}^{A}w_{ji}^{B}}{W_{tot}^{A}W_{tot}^{B}}
\label{rcrossprod}
\end{equation}
Although equation (\ref{rcrossprod}) is less precise to assess the reciprocity of every single layer compared to \ref{rmin}, it is a suitable measure of correlation across layers.
\noindent We can thereby measure correlation between flows in opposite direction ($reverse$ flows) and flows in the same direction ($synergic$ flows), for every couple of layers $A$ and $B$. In the former case, $r$ is a measure of the reciprocity for a multiplex composed by a couple of layers. In the latter case, $r$ is  no longer a measure of reciprocity because the layers' flows run in the same direction: we will name this measure $multiplexity$:

\begin{equation}
m_{Syn}^{AB}=\frac{\sum_{i\neq j}w_{ij}^{A}w_{ij}^{B}}{\sum_{i\neq j}w_{ij}^{A}\sum_{i\neq j}w_{ij}^{B}}
\label{multiplexity}
\end{equation}

It is noteworthy that equation (\ref{rcrossprod}) is scale free and therefore suitable for our multiplex, that displays different units and scale for each layer. It should be noted that the diagonal elements of the reciprocity matrix, contrary to those of the correlation matrix based on (\ref{pearson_syn}), aren't equal to $1$ by definition. This is due to the fact that equation \ref{rcrossprod} measures the reciprocity of the multiplex and, thus, scores $1$ on the diagonal only when the layer is symmetrical (fully reciprocated). 

\noindent Furthermore, reciprocity according to equation \ref{rcrossprod} is easily handled to compute null models, as the expected values of products are the products of expected values, for independent variables:

\begin{equation}
\langle W^{AB}_{\bigcap}\rangle=\sum_{i\neq j}\langle w_{ij}^{A} \rangle \langle w_{ji}^{B} \rangle
\end{equation}
Hence, a definition of correlation based on cross-products  reciprocity and integrated by corresponding null model (as previously shown for the binary case) is:
\begin{equation}
\rho^{AB}=\frac{r^{AB}-\langle r^{AB}\rangle}{1-\langle r^{AB}\rangle} 
\label{rhocrossprod}
\end{equation}

\noindent Therefore, the $\mu$ multiplexity between two layers is:
\begin{equation}
\mu^{AB}=\frac{m^{AB}-\langle m^{AB}\rangle}{1-\langle m^{AB}\rangle} 
\label{mu}
\end{equation}
where 
\begin{equation}
\langle r^{AB}\rangle=\frac{\sum_{i\neq j}\langle w_{ij}^{A}\rangle \langle w_{ji}^{B} \rangle}{\sum_{i\neq j} \langle w_{ij}^{A} \rangle \sum_{i\neq j} \langle w_{ji}^{B} \rangle}
\label{expr}
\end{equation}
and 
\begin{equation}
\langle m^{AB}\rangle=\frac{\sum_{i\neq j}\langle w_{ij}^{A}\rangle \langle w_{ij}^{B} \rangle}{\sum_{i\neq j} \langle w_{ij}^{A} \rangle \sum_{i\neq j} \langle w_{ij}^{B} \rangle}
\label{expm}
\end{equation}
\subsection*{Null models\label{sec_null}}
\begin{figure}
\includegraphics[width=0.48\textwidth]{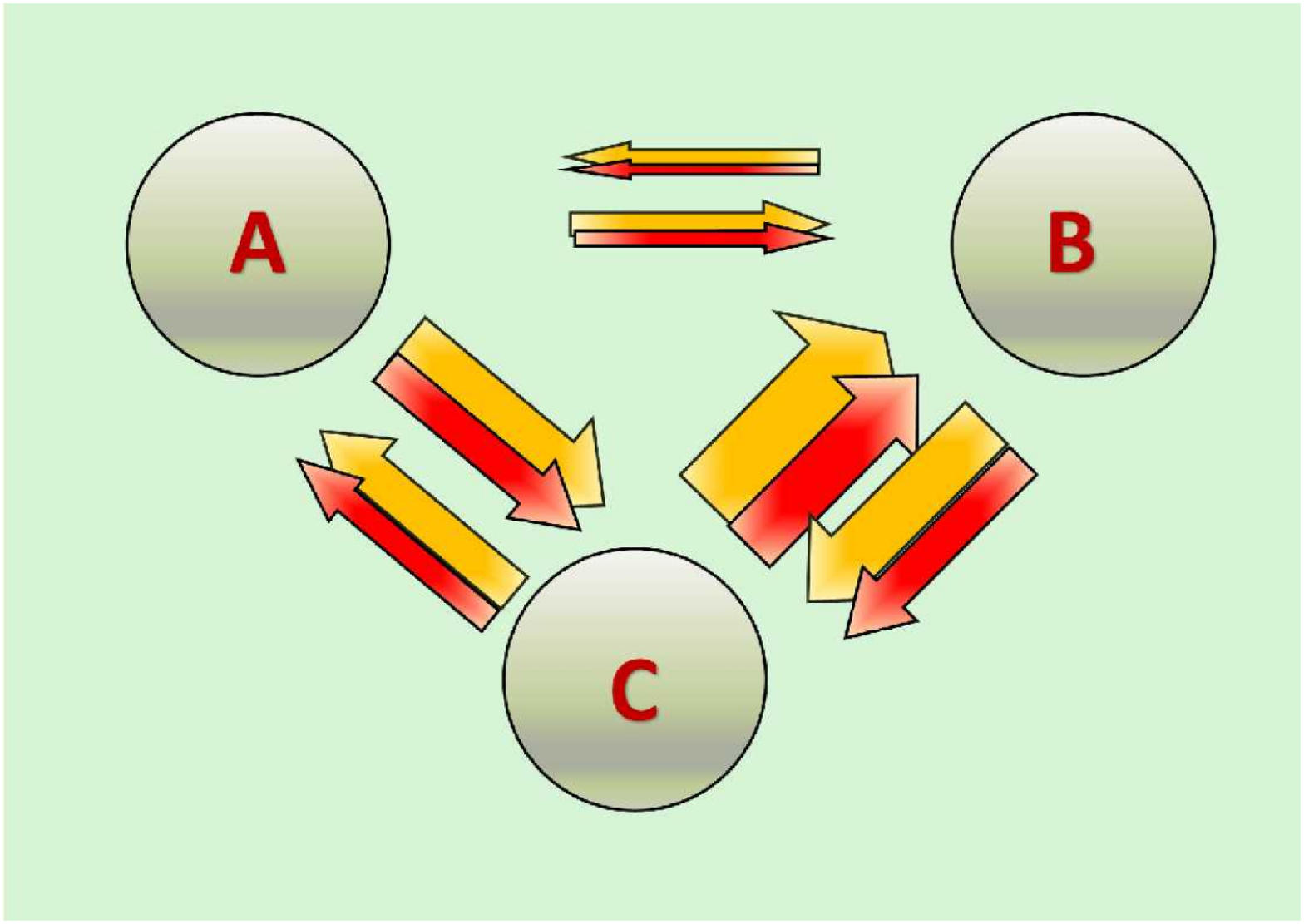}
\includegraphics[width=0.47\textwidth]{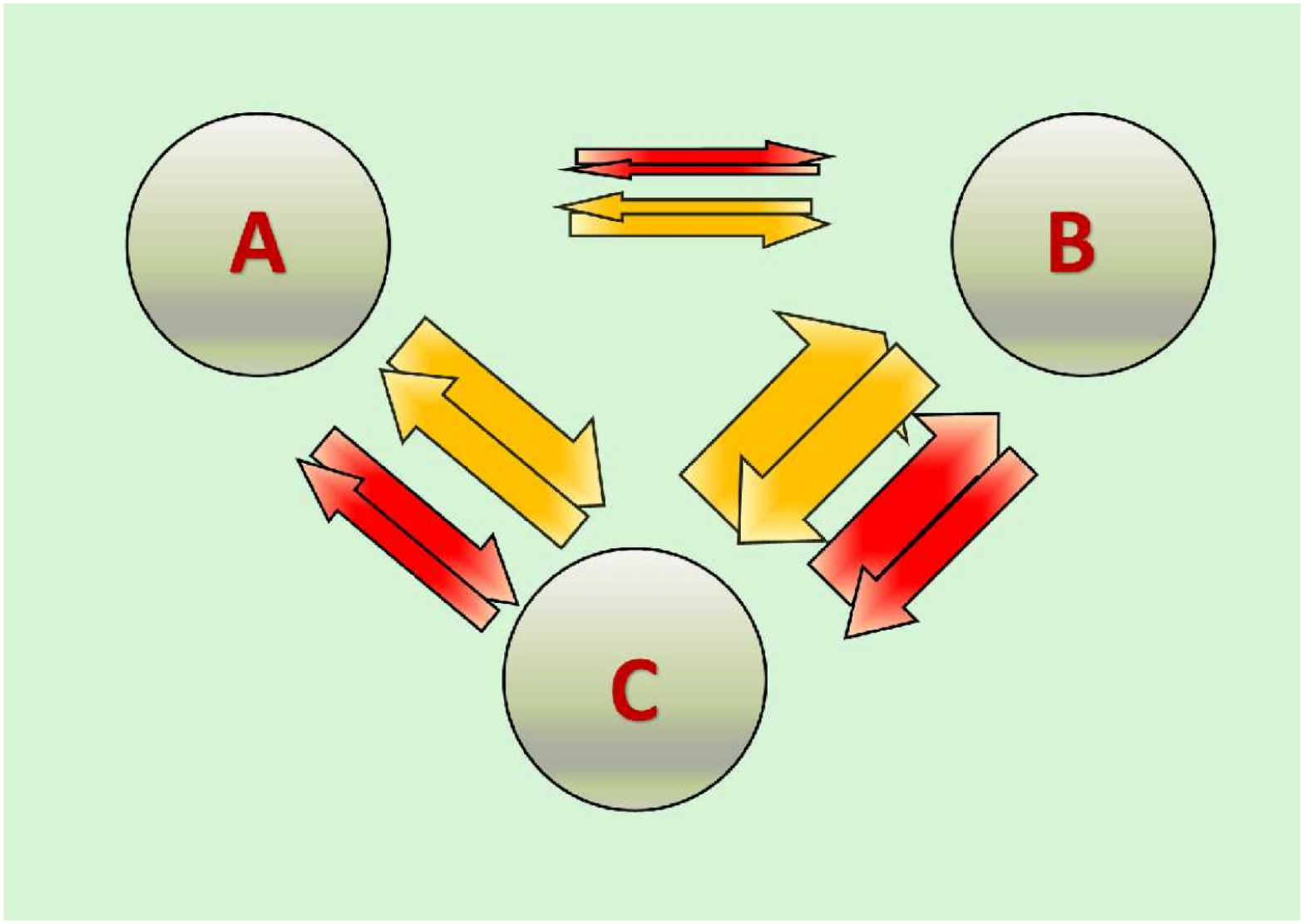}
\caption{Disentangling correlation and reciprocity across layers}
\label{S1 Fig}
\end{figure}

A null model is meant to test the statistics significance of results, on the one hand, and to enhance the correlation analysis by clearing them from undesired side effects generated by the complexity of the interactions' structure, on the other. Henceforth, the first step that needs to be done is that of formalising a $suitable$ null model for a correlation analysis of a multiplex based on a network reciprocity measure. Indeed, a null model has to preserve some features of the network that we know are relevant, but not informative for the ongoing investigation and what we suspect may affect our analysis.
Figure \ref{S1 Fig} shows, by the means of a very simple and stylized multiplex with three nodes and two layers, to what extent the reciprocity structure within each layer can affect the observed correlation between layers (width of the arrows indicates the intensity of the flows). \emph{Prime facie}, the red and the yellow layers seem to be very correlated when moving across nodes. Nevertheless, this is due the combined effect of the reciprocity level within each layer and the topology of the multiplex. For instance, node $C$ is a hub in both layers and thus node $A$ and $B$ tend to exchange with $C$ more than between each other. The entanglement of these two effects engenders an observed correlation between the two layers.

Interestingly, the degree of global symmetry of a network derives locally from the tendency of nodes to balance inflow and outflow \cite{mysymmetry2}. Overall the multiplex' layers tend to be weakly balanced, yet this effect seems to be stable in time. In other words, in both the financial and in environmental networks there are net exporter/importers of finance/matter in the OECD countries. To what extent this topology will affect our measures of correlation?

In order to clear correlation measures across layers of the topological effects, we adopt the formalism of Exponential Random Graphs or $p^*$ models, which allow to obtain maximally random ensembles of networks with specified constraints.
Exponential random graphs were first introduced in social network analysis \cite{WF,HL,snijders,pattison} and then recently rephrased within a maximum-entropy approach typical of statistical physics  \cite{newman_expo,snijders,pattison}.
Exponential Random Graphs are very useful when one needs to understand, as in our case, the expected effects of a given set of topological properties, $\vec{C}$ (such as the total weight, or the strength sequence) on the structure of networks.
Recently, a method based on the maximum-likelihood principle was proposed \cite{mymethod} in order to fit exponential random graphs to a real-world graph $\textbf{G}^*$ exactly \cite{mymethod}. This method provides null models which specify the effects of one or more constraints on the structure of the \emph{particular} network $\textbf{G}^*$, and hence allows to empirically detect patterns in the latter, identified as deviations from the model's predictions \cite{mymethod}. 
In this method, maximum-entropy exponential random graphs are generated by specifying an ensemble $\mathcal{G}$ of allowed graphs, and by looking for the probability $P(\textbf{G}|\vec{\theta})$ of generating a single graph $\textbf{G}$ in the ensemble in such a way that the Shannon entropy

\begin{equation}
S(\vec{\theta})\equiv-\sum_{\textbf{G}\in \mathcal{G}}P(\textbf{G}|\vec{\theta})\ln P(\textbf{G}|\vec{\theta})
\label{eq:entropy}
\end{equation}

\noindent is maximum, under the constraints that the probability is properly normalized, $\sum_{\textbf{G}\in\mathcal{G}}P(\textbf{G}|\vec{\theta})=1,\:\forall\vec{\theta}$, and that the expected value 

\begin{equation}
\langle \vec{C}\rangle_{\vec{\theta}}\equiv \sum_{\textbf{G}\in\mathcal{G}} \vec{C}(\textbf{G})P(\textbf{G}|\vec{\theta})
\end{equation}

\noindent of the set $\vec{C}$ of enforced topological properties equals the particular  value $\vec{C}^*\equiv \vec{C}(\textbf{G}^*)$ observed on the real network $\textbf{G}^*$:

\begin{equation}
\langle \vec{C}\rangle_{\vec{\theta}^*}=\vec{C}^*.
\label{eq:con}
\end{equation}

In the above expressions, $\vec{\theta}$ is a vector of Langrange multipliers allowing to tune the value of $\langle \vec{C}\rangle_{\vec{\theta}}$, and $\vec{\theta}^*$ is the specific value of $\vec{\theta}$ that makes $\langle \vec{C}\rangle_{\vec{\theta}}$ coincide with $\vec{C}^*$, as dictated by the maximum-likelihood principle \cite{mylikelihood}. The solution to the above constrained maximization problem is

\begin{equation}
P(\textbf{G}|\vec{\theta}^*)=\frac{e^{-H(\textbf{G}|\vec{\theta}^*)}}{Z(\vec{\theta}^*)}
\end{equation}

\noindent where

\begin{equation}
H(\textbf{G}|\vec{\theta}^*)=\vec{\theta}^*\cdot\vec{C}(\textbf{G})
\end{equation}

\noindent is sometimes called the \emph{graph Hamiltonian} and

\begin{equation}
Z(\vec{\theta}^*)=\sum_{\textbf{G}\in\mathcal{G}}{e^{-H(\textbf{G}|\vec{\theta}^*)}}
\end{equation}

\noindent is the \emph{partition function}, ensuring that the probability is properly normalized. The above formal results translate into specific quantitative expectations when a particular choice of the constraints, $\vec{C}$, is made. 

Once the numerical values of the Lagrange multipliers are found, they can be used to find the ensemble average, $\langle X\rangle^*$, of any topological property $X$ of interest:

\begin{equation}
\langle X\rangle^*=\sum_{\mathbf{G}\in\mathcal{G}}X(\mathbf{G})P(\mathbf{G}|\vec{\theta}^*).
\end{equation}

The exact computation of the expected values can be very difficult. For this reason it is often necessary to rest on the linear approximation method even if, in what follows, the only approximation will be that of treating the expected value of a ratio, as the ratio of the expected values: $\langle n/d\rangle\simeq \langle n\rangle/\langle d\rangle$.

The next subsections will be devoted to the description of the null model appropriate for our analysis.

\subsection*{The Weighted Reciprocated Configuration Model (WRCM)\label{sec_wrho}}

The financial networks are weakly reciprocated, whereas environmental networks are significantly reciprocated. As it was previously highlighted, the reciprocity structure of every layer will affect the correlation analysis of the multiplex. Therefore, it is of prominent interest for the present analysis developing a null model that incorporates both structures to filter out the combined effect of the first degree topology (import, export sequence) and the reciprocity level of every single layer. We want a null model that constrains for every node the total import, total export and the share that is mutually exchanged (reciprocated)  \cite{Wrecip}. The graph Hamiltonian thus becomes:

\begin{equation}
H(\mathbf{G}|\vec{\theta})=\sum_{i}(\alpha_{i}s_{i}^{\rightarrow}+\beta_{i}s_{i}^{\leftarrow}+\gamma_{i} s_{i}^{\leftrightarrow})
\end{equation}

\noindent where, now, $\vec{\theta}\equiv\{\vec{\alpha},\:\vec{\beta},\:\vec{\gamma}\}$ and

\begin{eqnarray}
s_{i}^{\rightarrow}\equiv \sum_{j(\neq i)}w_{ij}^{\rightarrow},\:s_{i}^{\leftarrow}\equiv \sum_{j(\neq i)}w_{ij}^{\leftarrow},\:s_{i}^{\leftrightarrow}\equiv \sum_{j(\neq i)}w_{ij}^{\leftrightarrow}
\end{eqnarray}

\noindent with obvious meaning of the symbols (defined above). The partition function now becomes

\begin{eqnarray}
Z(\vec{\theta})&=&\prod_{i<j}\frac{(1-x_{i}x_{j}y_{i}y_{j})}{(1-x_{i}y_{j})(1-x_{j}y_{i})(1-z_{i}z_{j})}\equiv\nonumber\\
&\equiv&\prod_{i<j}Z_{ij}^{WRCM}(\vec{\theta})
\end{eqnarray}

\noindent and the likelihood is

\begin{eqnarray}
\ln P(\textbf{G}^*|\vec{\theta})&=&\sum_{i< j}[(w_{ij}^{\rightarrow})^*\ln(x_{i}y_{j})+(w_{ij}^{\leftarrow})^*\ln(x_{j}y_{i})+\nonumber\\
&+&(w_{ij}^{\leftrightarrow})^* \ln(z_{i}z_{j})-\ln Z_{ij}^{WRCM}(\vec{\theta})].
\end{eqnarray}

The solution to this optimization problem, with respect to $\vec{x}$, $\vec{y}$ and $\vec{z}$, can be found by solving the following system:

\begin{eqnarray}
\left\{ \begin{array}{ll}
s^{\rightarrow}_i(\textbf{G}^*) &= \sum_{j\ne i}\langle w_{ij}^{\rightarrow}\rangle_{\vec{\theta}^*}=\langle s_{i}^{\rightarrow}\rangle_{\vec{\theta}^*},\quad \forall i\\
s^{\leftarrow}_i(\textbf{G}^*) &=\sum_{j\ne i}\langle w_{ij}^{\leftarrow}\rangle_{\vec{\theta}^*}=\langle s_{i}^{\leftarrow}\rangle_{\vec{\theta}^*},\quad \forall i\\
s_{i}^{\leftrightarrow}(\textbf{G}^*) &= \sum_{j\ne i}\langle w_{ij}^{\leftrightarrow}\rangle_{\vec{\theta}^*} = \langle s_{i}^{\leftrightarrow}\rangle_{\vec{\theta}^*},\quad \forall i
\end{array} \right.
\label{eq:wrcm}
\end{eqnarray}

\noindent where

\begin{equation}
\langle w_{ij}^{\rightarrow}\rangle_{\vec{\theta}^*}=\frac{x^*_iy^*_j(1-x^*_jy^*_i)}{(1-x^*_iy^*_j)(1-x^*_ix^*_jy^*_iy^*_j)},
\end{equation}
\begin{equation}
\langle w_{ij}^{\leftarrow}\rangle_{\vec{\theta}^*}=\frac{x^*_jy^*_i(1-x^*_iy^*_j)}{(1-x^*_jy^*_i)(1-x^*_ix^*_jy^*_iy^*_j)},
\end{equation}
\begin{equation}
\langle w_{ij}^{\leftrightarrow}\rangle_{\vec{\theta}^*}=\frac{z^*_iz^*_j}{1-z^*_iz^*_j}.
\end{equation}

By the definition of the WRCM model, we not only recover the result that the global reciprocity is equal to the observed one (implying $r\equiv \langle r\rangle_{WRCM}$ and $\rho_{WRCM}\equiv0$, also valid for the WRM). Now, all the vertex-level, strength sequences are exactly reproduced, implying that the reciprocity is reproduced at a \emph{local} level.
%%%%%%%%%%%%%%%%%%%%%

\begin{figure}
\begin{center}
\includegraphics[width=0.9\textwidth]{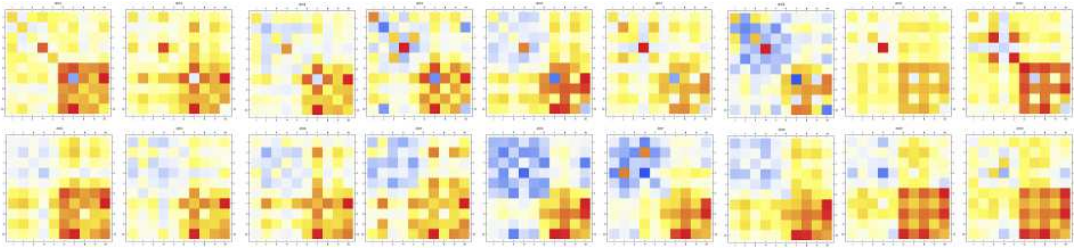}
\end{center}
 \caption{Time evolution of the $\mu$ multiplexity correlation matrix (synergic flows), first row, and $\rho$ reciprocity correlation matrix (reverese flows), second row,2002-2010}
\label{S2 Fig}
\end{figure}

In what follows, we are able to filter out first order and second order topological effects, generated by a single-layer topology, from cross-layers correlations. The enhanced measures of synergic and reverse correlations will thus be: $\rho$-correlations (eq. \ref{rhocrossprod}) and  $\mu$-correlations (eq. \ref{mu}).
Figure \ref{S2 Fig} shows the correlation matrix for synergic flows and reverse flows for the entire period of investigation (temperature maps). 
Shades of yellow indicate a positive correlation and shades of blue negative ones, from lighter to more intense for stronger correlations.
According to our analysis, the correlation between synergic environmental and financial flows is weakly positive, but persistent over time. In contrast, the correlation between reverse environmental and financial flows is less stable across time and it does not highlight a clear pattern. 
- environmental, both synergic and reverse, flows are highly and positively correlated within themselves. 
- correlation across financial synergic flows is volatile and varies from weakly positive to negative, but reverse correlation is mostly negative, despite its variation in intensity.

\subsection*{Local reciprocity and local multiplexity\label{sec_local_rho}}
We have defined a measure of layers' reciprocity (equation \ref{rcrossprod}) and multiplexity and upon that a measure of  layers correlation for synergic and reverse flows of the multiplex (equations \ref{rhocrossprod} and \ref{mu}). This latter measure of correlation embodies a null model, the RWCM (equation \ref{eq:wrcm}) that preserves the topology and the reciprocity structure of each layer. Therefore, this measure enables us to observe cross correlation between layers that are not trivially produced by the overlapping of layers with different reciprocity structure. 
We can further extrapolate a measure of local reciprocity normalized on every column (export)  for every couple of country $i$ and $j$ and pair of layers $A$ and $B$:
\begin{equation}
r_{ij}^{AB}=\frac{w_{ij}^{A}w_{ji}^{B}}{\sum_{j}w_{ij}^{A}\sum_{j}w_{ji}^{B}}
\label{localr}
\end{equation}
and a corresponding local multiplexity:
\begin{equation}
m_{ij}^{AB}=\frac{w_{ij}^{A}w_{ij}^{B}}{\sum_{j}w_{ij}^{A}\sum_{j}w_{ij}^{B}}
\label{localm}
\end{equation}

In what follows, equation (\ref{localr}) enables us to compute a matrix of local, cross-layer reciprocities for a couple of layers and a couple nodes, and to use them as measures of spurious (local) correlations.
This measures score $1$ when all links' weights of nodes $i$ and $j$, for both layers $A$ and $B$, are placed between node $i$ and $j$.
Likewise global reciprocity across layers (equations \ref{rcrossprod} and \ref{multiplexity}) can be enhanced by incorporating the null model in the $\rho$ and $\mu$ correlations, local reciprocity measurements can be equally improved with the use of the null model. Therefore, local $\rho$ and $\mu$  for a couple of nodes $i$ and $j$, between layer $A$ and $B$ are:
\begin{equation}
\mu_{ij}^{AB}=\frac{m_{ij}^{A}-\langle m_{ij}^{AB}\rangle}{1-\langle m_{ij}^{AB}\rangle}
\label{localmu}
\end{equation}

\begin{equation}
\rho_{ij}^{AB}=\frac{r_{ij}^{A}-\langle r_{ij}^{AB}\rangle}{1-\langle r_{ij}^{AB}\rangle}
\label{localrho}
\end{equation}
In section \ref{sec_error} it will be explained why negative values of $\rho$ and $\mu$ correlations are biased (RWCM tends to overestimate negative correlations and underestimate positive correlations). The null model, therefore, set a \emph{threshold} for significant positive correlations across layers and between countries in the multiplex which is informative, as it remove the effects of the single-layer topology and reciprocity.
In what follows, we will proceed considering positive $\mu$ and $\rho$ correlations as significantly signed by the null model. 
In Figure $1-e$ of the article, yellow dots signal positive correlations whereas blue dots negative ones. Yellow lines at the far right show that some countries with a high financialisation (last columns) exchange finance with environment, prominently with countries with low financialisation (first columns). This correlation is expressed by yellow dots in the top-right block of the $\mu$ and $\rho$  matrix. Conversely, countries with low finacialisation exchange mainly with countries with high finacialisation, as the yellow dots in the bottom-left block of the matrix indicates.
We have replicated this exercise for all cross-layer correlation between financial and environmental layers and we have found the existence of a common topology among the couples of financial-environmental layers: positive correlation between each couple of outward financial and inward environmental (either synergic or reverse) flows strengthens as the level of financialization increases. In the next section, we will also investigate whether reverse correlations are more significant than synergic correlations and whether this pattern is more accentuated in reverse than in synergic correlation matrixes.
All the cross-layer correlations between financial layers and environmental layers display the same distribution of yellow dots, signalling the existence of a common topological pattern in all the couple of layers.
The analysis of the cross-layer correlation between the four financial layers and the five environmental layers in aggregate highlights the pivotal roles of five countries among the others. These five countries are hubs in the topology of the flows between finance and environment: UK, Germany, Bel-Lux, France and USA. This peculiar topology will be addressed in the next section. 

\subsection*{Correlation between layers}
Table \ref{tab3} shows the average Pearson correlation index for synergic and reverse and standard deviation. Table \ref{tab4} shows the average reciprocity and multiplexity respectively. Interestingly, reverse correlations between financial layers and environmental layers are generally stronger and more frequent than synergic ones. Reverse correlations between the environment and finance are also more stable in time. 
A second interesting result is the dominance of the equity market. Equities is the  financial layer most correlated to environmental layers, followed by total debts securities (TD)and short term debts securities (SD). It is noteworthy that equities and debts' securities are generally considered speculative financial tools. Surprisingly, foreign direct investments (FDI), that accounts for direct, long term investments in the economy, is generally weakly correlated to the environment.

The most correlated environmental layer to the financial layers is $SO_{2}$ and the second most correlated is $NO_{x}$. The production of both is generally associated with combustion and prominently with \emph{inefficient} combustion, hinting at a specific, antiquate electricity mix, heavy industry, such, metals, metallurgy or machinery and chemical industry. However, more research is needed to clarify such links.

\begin{table}[h]
\caption{Average Pearson correlation of the financial sector (layers 1-5) with the environment (layers 6-10)}
\begin{tabular}{llllll}
\hline
 Layer &$NO_{x}$& $PM10$& $SO_{2}$ &$CO_{2}$& Water\\
\hline
FDI syn&$0.25\pm0.03$ &$0.10\pm0.03$ &$0.28\pm0.03$ & $0.18\pm0.04$&$0.12\pm0.06$ \\
FDI rev&$0.23\pm0.05$ &$0.11\pm0.06$ &$0.28\pm0.05$ & $0.14\pm0.06$&$0.25\pm0.07$ \\
Equity syn&$0.32\pm0.09$ &$0.16\pm0.06$ &$0.34\pm0.09$ & $0.24\pm0.12$& $0.27\pm0.17$ \\
Equity rev&$0.34\pm0.15$ &$0.15\pm0.09$ &$0.36\pm0.13$ & $0.23\pm0.13$& $0.26\pm0.10$ \\
SD syn&$0.25\pm0.05$ &  $0.08\pm0.03$  &  $0.27\pm0.04$ &  $0.17\pm0.02$ &$0.11\pm0.05$ \\
SD rev&$0.24\pm0.05$ &  $0.12\pm0.05$  &  $0.28\pm0.06$ &  $0.15\pm0.05$ &$0.26\pm0.09$ \\
LD syn&$0.15\pm0.03$ &  $0.05\pm0.02$  &  $0.17\pm0.02$&  $0.09\pm0.02$ & $0.12\pm0.04$ \\
LD rev&$0.16\pm0.05$ &  $0.05\pm0.03$  &  $0.16\pm0.05$&  $0.10\pm0.05$ & $0.10\pm0.04$ \\
TD syn&$0.32\pm0.09$ & $0.13\pm0.04$ &$0.34\pm0.08$ &$0.23\pm0.08$ & $0.22\pm0.15$ \\ 
TD rev&$0.33\pm0.12$ & $0.16\pm0.08$ &$0.37\pm0.10$ &$0.23\pm0.13$ & $0.29\pm0.10$ \\
\hline
\end{tabular}

\label{tab3}
\end{table}
\begin{table}[h]
\caption{Average reciprocity and multiplexity of the financial sector with the environment, values rescaled to $10^{2}$}
\begin{tabular}{llllll}
\hline
 Layer &$NO_{x}$& $PM10$& $SO_{2}$ &$CO_{2}$& Water\\
\hline
FDI mul&$0.45\pm0.06$ &$0.30\pm0.05$ &$0.39\pm0.06$ & $0.39\pm0.07$&$0.32\pm0.09$ \\
FDI rec&$0.42\pm0.07$ &$0.33\pm0.09$ &$0.45\pm0.07$ & $0.32\pm0.09$&$0.56\pm0.13$ \\
Equity mul&$0.63\pm0.22$ &$0.50\pm0.17$ &$0.62\pm0.20$ & $0.55\pm0.18$&$0.73\pm0.52$ \\
Equity rec&$0.67\pm0.35$ &$0.50\pm0.31$ &$0.66\pm0.29$ & $0.56\pm0.39$&$0.67\pm0.21$ \\
SD mul&$0.45\pm0.09$ &$0.29\pm0.07$ &$0.47\pm0.07$ & $0.38\pm0.09$&$0.31\pm0.08$ \\
SD rec&$0.45\pm0.08$ &$0.36\pm0.10$ &$0.48\pm0.08$ & $0.34\pm0.07$&$0.62\pm0.21$ \\
LD mul&$0.43\pm0.09$ &$0.26\pm0.09$ &$0.44\pm0.06$ & $0.32\pm0.07$&$0.46\pm0.13$ \\
LD rec&$0.44\pm0.11$ &$0.24\pm0.09$ &$0.42\pm0.09$ & $0.34\pm0.11$&$0.38\pm0.13$ \\
TD mul&$0.55\pm0.15$ &$0.39\pm0.11$ &$0.55\pm0.13$ & $0.47\pm0.14$&$0.53\pm0.33$ \\
TD rec&$0.57\pm0.20$ &$0.45\pm0.18$ &$0.58\pm0.16$ & $0.47\pm0.24$&$0.67\pm0.21$ \\ 
\hline
\end{tabular}
\label{tab4}
\end{table}

\subsection*{Correlation between countries}
In section \ref{sec_local_rho} we have proposed a methodology to investigate correlations between countries and between layers and weave shown that these measures can be implemented with the adopted null model. 
However, as was previously highlighted, only positive correlations are truly informative. In what follows, we can translate the correlation matrix depicted in Figure $1-e$ into a binary, directed matrix, by replacing positive correlations with $1$ and negative or zero-correlations with $0$. The binary directed matrix can be further manipulated in order to extract a binary undirected matrix, according to the relationship: $U=A*(A^{T})$, where $A$ is the binary directed matrix and $U$ is the undirected matrix. The largest connected component of $U$ is the backbone of the correlation network between two layers. Figure \ref{fig3} shows the backbone for the financial-environmental multiplex. Hubs of the backbone are Germany (27), Bel-Lux (29), USA (33) and France (32) for synergic flows and UK (25), Germany and Bel-Lux for reverse flows. The analysis of the backbones of all the 25 pair of financial-environmental layers  shows that nodes 24, 26, 30 and 31 are generally poorly connected. That is to say that, despite being highly financialized countries, Canada, Nedherlands, Australia and Israel are marginal in the correlation network of financial-environmental flows. This not to say that they \emph{don't exchange} either finance or environment, or both, but that are less strongly coupled to the activities of other OECD countries.

\begin{figure}
\begin{center}
\includegraphics[width=0.9\textwidth]{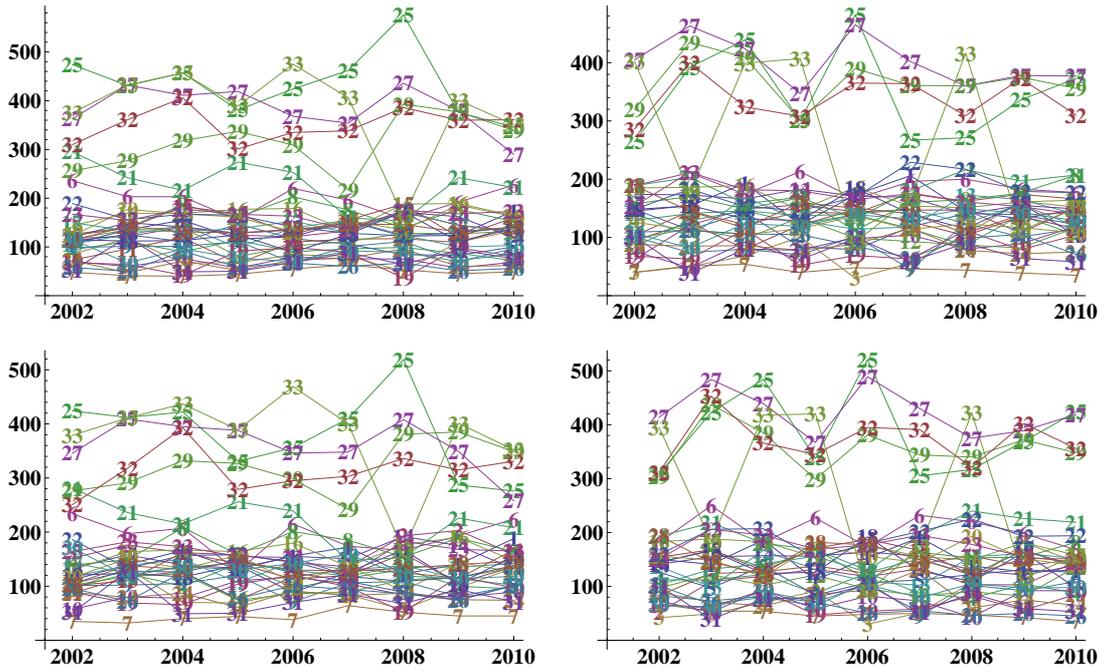}
\end{center}
    \caption{Time evolution of number of the number of incoming (left) and (right) outgoing links for synergic flows ($\mu$ correlations) and the second row for reverse flows ($\rho$ correlations).}
\label{S3 Fig}
\end{figure}

We identified the number of positive links for each country and plotted on a temporal axis to further extended the binary analysis of the topology of significant correlations to the whole multiplex, focusing on correlations between financial layers on the one hand and environmental layers on the other. In Figure \ref{S3 Fig} shows the time evolution of number of directional links for all OECD countries between all the couple of financial and environmental layers.  The first row of Figure  \ref{S3 Fig} shows the number of incoming and outgoing (directional) links for synergic flows ($\mu$ correlations) and the second row for reverse flows ($\rho$ correlations). Correlations always report a financial layer versus an environmental layer, thus, an outgoing link means an \emph{export in finance positively correlated to an import in environment}. According to our analysis, five countries among the OECD group stick out in terms of number of links: USA (33), France (32),  Bel-Lux (29), Germany (27) and UK (25). Therefore, we must conclude that there is a peculiar, common topology in all the cross-correlation networks, featured by the same five central nodes. Furthermore,it is worth noting that this topology is stable over time and robust to shocks, as it seems to have overcome the crisis of 2008 almost unaltered (with the exception of USA).

Indeed, these five countries are pivotal in the financial system and in the network of trades (ITN). A previous analysis of the correlations between financial layers and trades emphasize their leading role \cite{andreas1,andreas2}. 

\subsection*{Node balancing and layers correlation}
\begin{figure}
\begin{center}
\includegraphics[width=0.45\textwidth]{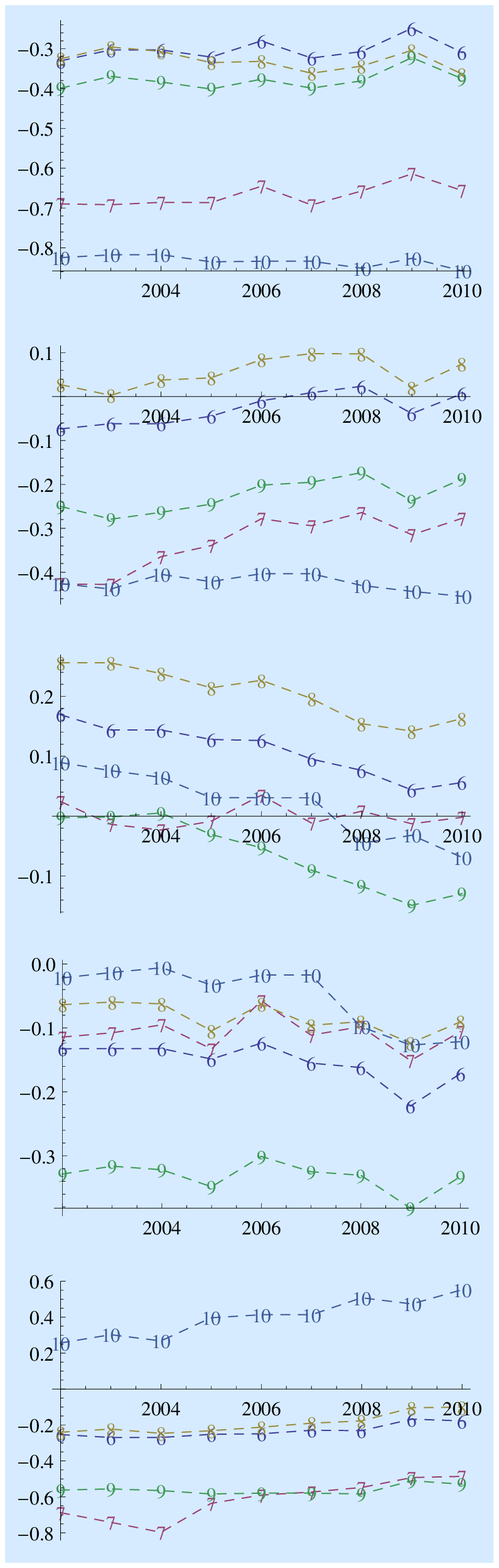}
\includegraphics[width=0.45\textwidth]{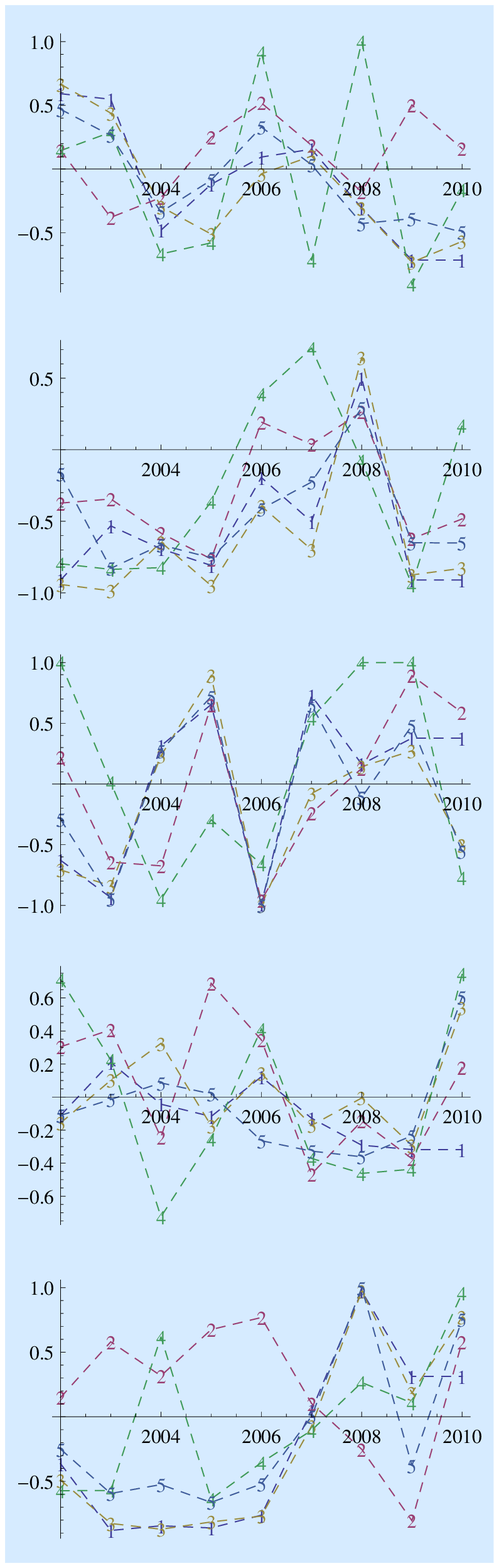}
\end{center}
 \caption{Time evolution of the imbalance of 5 financial layers (layers 1-5), panel a) and the 5 environmental layers (layers 6-10), panel b), for the five hubs (from top to the bottom): UK, Germany, Bel-Lux, France, USA.}
\label{S4 Fig}
\end{figure}

Albeit the filtering of the effects of topology and reciprocity of every single layer on the correlation between layers, node balancing (the balance between exports and imports) might still be crucial in explaining the observed patterns. The ten layers are generally unbalanced, meaning that nodes tend to be either net importers or exporters of finance or environment. Are those five central nodes in the correlation networks also hubs in the financial system and/or in the productive space? Are they net exporters or importers respectively?
Predictably, these five countries are net importers of environment, with some exceptions, like USA in the water flows, Germany in PM10 and Bel-Lux in most environmental layers, albeit with a decreasing trend. It is worth noting that these five countries are also net importers of finance. Furthermore, nodes' imbalance of the financial layers seems to be more unstable compared to the environmental layers, but this is not a surprise given the volatility of the financial sector compared to the real economy (Figure \ref{S4 Fig}) . Despite this volatility, correlations, and prominently reverse correlations, between financial layers and environmental layers are stable along time (see Table \ref{tab3}), hinting to some broader, fundamental process underpinning the economy and linking trades to finance.

\begin{figure}
\begin{center}
\includegraphics[width=0.45\textwidth]{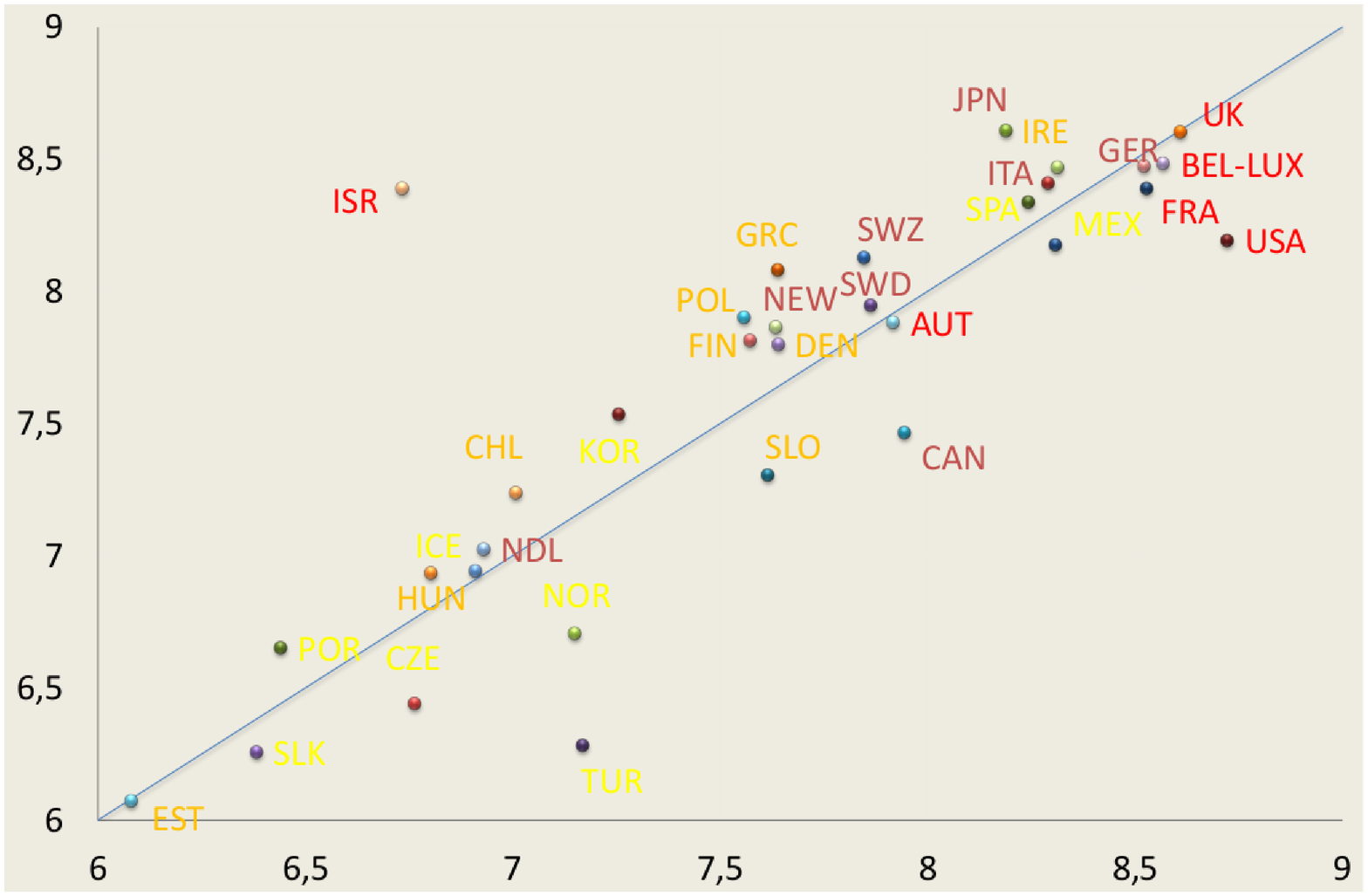}
\includegraphics[width=0.45\textwidth]{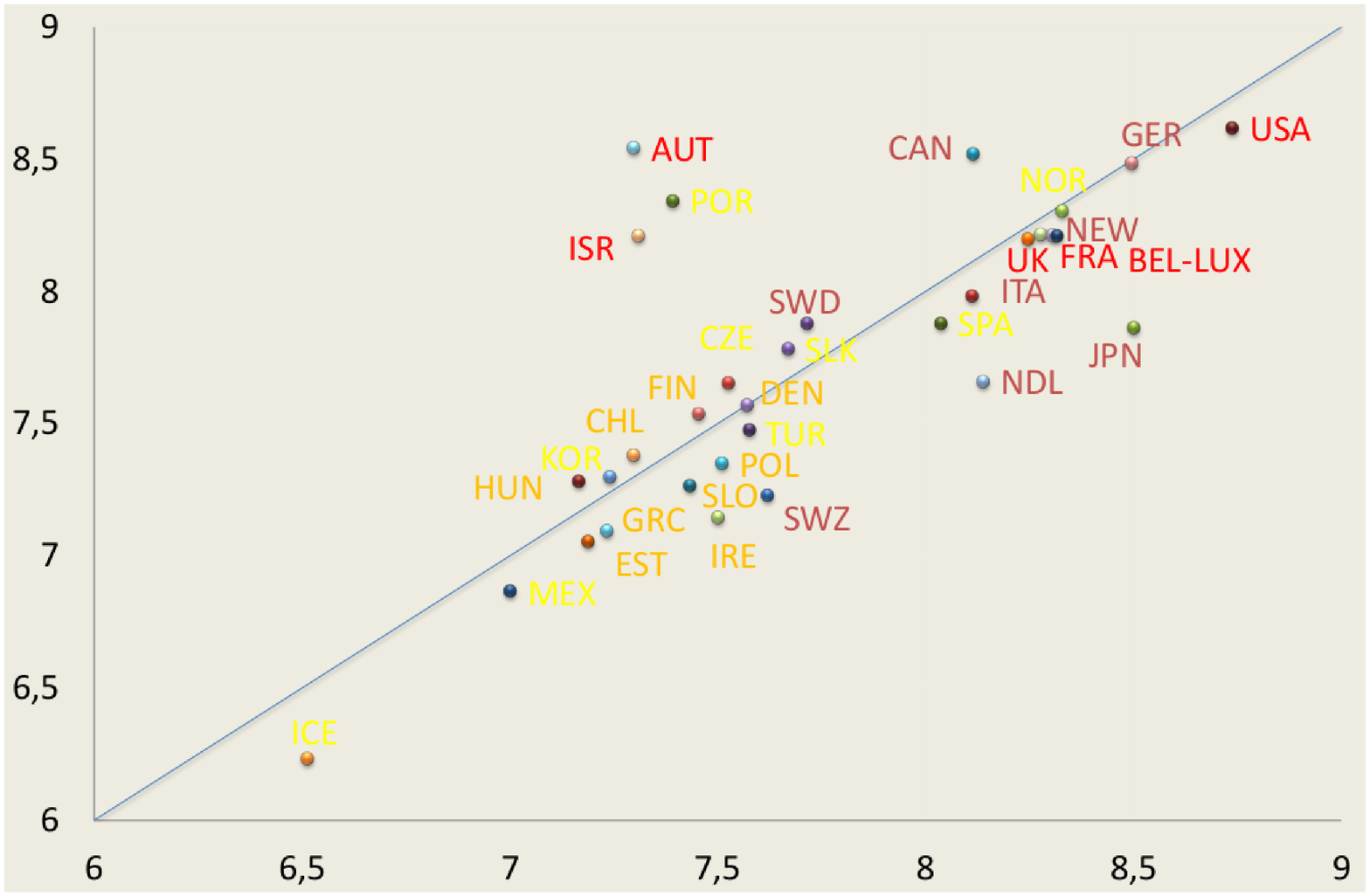}
\end{center}
    \caption{Financial imbalance, (Exp-Imp, thousand dollars, layers 1-4) for the 33 OECD countries (average 2002-2010). Colours from light yellow to bright red indicate increasing financialisation. In the region above the line exports exceeds imports. Values are reported on log-10-scale}
\label{S5 fig}
\end{figure}

Figure \ref{S5 fig} show the Export ($y$ axis) and the Imports ($x$ axis) on a $Log_{10}$ scale for the five financial layers and for trade in mass units. Countries are labelled with four colors indicating four degree of finanzialization, from light yellow to red. Countries' position along the principal axis shows the magnitude of their economy. OECD countries are broadly distributed along the line of balance and regardless of the size of their economy or their financialisation, they divide themselves in net-exporter or net-importer,  showing an inner, global balance, of mass and money.
However, contrary to the five hubs, most of OECD countries are alternatively net importers or finance or net importers of mass and environmental load. Furthermore, countries that are net importer of environment tend to be net importer in all the environmental networks, and \emph{vice versa}. Not the same can be said for financial layers: we do not generally witness to a position of net importers/exporters in all the layers simultaneously.
It is also interesting to note that not all the least financialised countries are net importer of finance/exporter of environment likewise not all the most financialised countries are net exporter of finance/importer of environment.
Furthermore, countries that are net importer/exporter of finance can be net importer/exporter of environment too, like in the case of the five hubs. The two systems indeed do not seem to level off each other.

\subsection*{Error analysis\label{sec_error}}
The present analysis aims at investigating the durable and stable relationships between financial and environmental layers in OECD economies. Therefore, we trace correlations on the average values of trade flows along time. Hence, a first source of error of our analysis is the time variation of the analysis. Financial layers are much more volatile compared to environmental layers, both in terms of total volumes and local balancing of import/export. Correlation measures involving equities (layer 2) might be significantly affected by the year 2008, which saw a dramatic drop of money in the equity market diverted to other assets. We tested correlations between 2002-2010 without the year 2008. Despite decreasing the correlation of equities to the remaining layers, overall results  hold and equities along with TD are still considerably more correlated to the environment than the other financial layers.

\begin{figure}
\begin{center}
\includegraphics[width=0.9\textwidth]{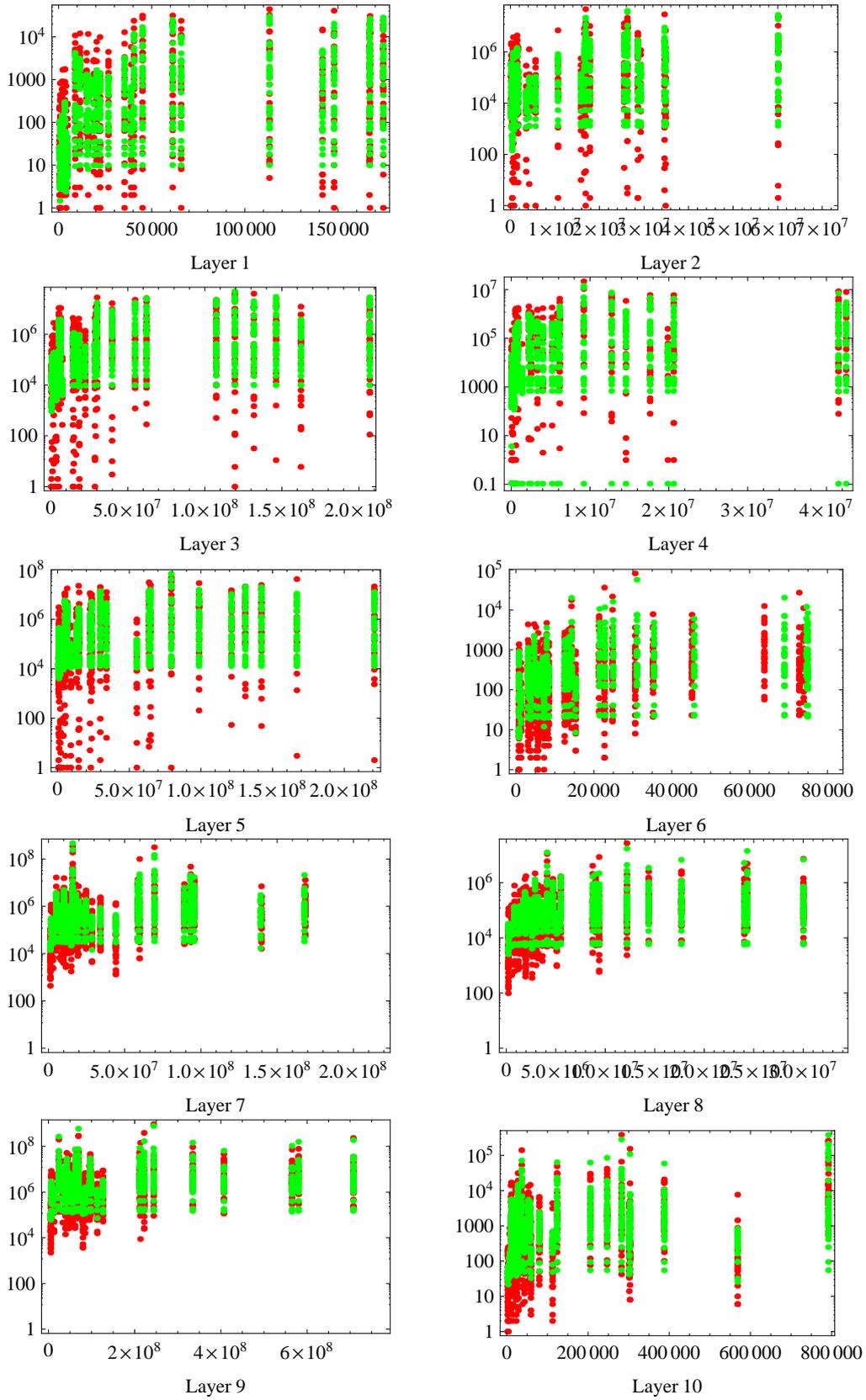}
\end{center}
 \caption{RWCM null model: plot of expected (green) vs observed (red) links weight over the nodes strength (export)}
\label{S6 Fig}
\end{figure}

We further assess the sensitiveness of Pearson correlations to time variation within layers by jack-knifing the time sample and evaluating the error magnitude (the variance of the jack-knife estimator). According to our error analysis, the ranking of correlations expressed in the analysis are significant. 
It also interesting that reverse correlations are in general more stable then synergic correlations, by one order of magnitude. 
As previously highlighted, our analysis, when endowing the null model, takes into account just positive correlations. This is because the null model described in section \ref{sec_null} tends to overestimate negative correlation. This effect is due to the log-likelihood maximization that delivers non-negative values. Since, the financial layers are sparse, with an average density (connectance) of $0.5$, the correlation index $\rho$ computed for every couple of nodes is negative any time a weight $w_{ij}$ for either layer $A$ or $B$ is null, and thus $r_{ij}$ will be zero (whereas the expected value of $r_{ij}$ is always positive).
Furthermore, as it is clear form Figure \ref{S6 Fig}, the null model tends to overestimate links' weights. Hence, positive correlations are \emph{conservative} estimations of correlation: positive correlations that survive the combination of those two effects are thus more informative and trustworthy.
In what follows, differently to Table \ref{tab3} and \ref{tab4}, average values are computed separately on the observed and expected matrices from 2002 to 2010 and successively $\mu$ and $\rho$ are calculated. In Table \ref{tab5}  we show results of the $\mu$ and $\rho$ (positive) correlations (equations \ref{rhocrossprod} and \ref{mu}) for the five financial layers compared with the five environment layers. Missing values refers to negative correlations and are omitted. 
Values are averaged over time (9 years) and over the environmental layers. Results in Table \ref{tab3}
generally hold under the scrutiny of the null model.
The most correlated financial layers are Equities, TD and lastly SD. Reverse flows are generally more correlated to the environment than synergic flows, except for the FDI layer where synergic flows are more correlated than reverse flows for all the five environmental layers and TD, that displays higher $\mu$ correlations with three layers out of with.

\begin{table}[h]
\caption{Average Pearson correlation of the financial sector(layers 1-5) with the environment (layers 6-10), years 2002-2010, without 2008}
\begin{tabular}{llllll}
\hline
 Layer &$NO_{x}$& $PM10$& $SO_{2}$ &$CO_{2}$& Water\\
\hline
FDI syn&$0.26\pm0.04$ &$0.10\pm0.03$ &$0.29\pm0.03$ & $0.19\pm0.03$&$0.12\pm0.06$ \\
FDI rev&$0.23\pm0.06$ &$0.11\pm0.06$ &$0.27\pm0.06$ & $0.13\pm0.06$&$0.26\pm0.06$ \\
Equity syn&$0.32\pm0.10$ &$0.17\pm0.06$ &$0.34\pm0.10$ & $0.26\pm0.10$& $0.26\pm0.18$ \\
Equity rev&$0.33\pm0.15$ &$0.14\pm0.10$ &$0.34\pm0.13$ & $0.21\pm0.16$& $0.27\pm0.10$ \\
SD syn&$0.25\pm0.05$ &  $0.09\pm0.04$  &  $0.28\pm0.04$ &  $0.18\pm0.05$ &$0.11\pm0.05$ \\
SD rev&$0.24\pm0.05$ &  $0.12\pm0.05$  &  $0.28\pm0.06$ &  $0.15\pm0.05$ &$0.26\pm0.09$ \\
LD syn&$0.16\pm0.03$ &  $0.05\pm0.02$  &  $0.17\pm0.03$&  $0.10\pm0.02$ & $0.12\pm0.03$ \\
LD rev&$0.15\pm0.06$ &  $0.04\pm0.03$  &  $0.15\pm0.05$&  $0.09\pm0.04$ & $0.10\pm0.05$ \\
TD syn&$0.33\pm0.09$ & $0.14\pm0.05$ &$0.35\pm0.09$ &$0.25\pm0.07$ & $0.21\pm0.16$ \\ 
TD rev&$0.33\pm0.12$ & $0.15\pm0.08$ &$0.36\pm0.11$ &$0.21\pm0.13$ & $0.32\pm0.09$ \\
\hline
\end{tabular}
\label{tab5}
\end{table}

\newpage
\begin{small}

\end{small}
\end{document}